\documentclass[10pt]{article}
\usepackage[OE]{express}
\usepackage{epstopdf}
\usepackage[labelformat=simple]{subcaption}

\usepackage{soul}
\usepackage{multirow}
\usepackage[margin=0.5 in]{caption}
\begin{document}
\title{Signaling on the Continuous Spectrum of Nonlinear Optical Fiber}

\author{Iman Tavakkolnia\authormark{*} and Majid Safari}
\address{Institute for Digital Communications, School of Engineering, University of Edinburgh, EH9 3JL, Edinburgh, UK }
\email{\authormark{*}i.tavakkolnia@ed.ac.uk}	
	

\begin{abstract}
This paper studies different signaling techniques on the continuous spectrum (CS) of nonlinear optical fiber defined by nonlinear Fourier transform. Three different signaling techniques are proposed and analyzed based on the statistics of the noise added to CS after propagation along the nonlinear optical fiber. The proposed methods are compared in terms of error performance, distance reach, and complexity. Furthermore, the effect of chromatic dispersion on the data rate and noise in nonlinear spectral domain is investigated. It is demonstrated that, for a given sequence of CS symbols, an optimal bandwidth (or symbol rate) can be determined so that the temporal duration of the propagated signal at the end of the fiber is minimized. In effect, the required guard interval between the subsequently transmitted data packets in time is minimized and the effective data rate is significantly enhanced. Moreover, by selecting the proper signaling method and design criteria a distance reach of 7100 km is reported by only singling on CS at a rate of 9.6 Gbps.
\end{abstract}




\section{Introduction}

Extensive efforts have been made to develop new techniques for higher achievable data rates in optical fiber communication particularly over long-haul systems, where nonlinear effects are significant \cite{Cartledge,capacity,winzer}. Systems based on the nonlinear Fourier transform (NFT) have recently attracted attention among researchers as an approach to overcome fiber nonlinearity and to achieve the true capacity of optical fiber links \cite{yousefi1,yousefi2,yousefi3,Turitsynoptica}. In the absence of any perturbation (e.g., amplifier noise), the nonlinear channel defined by nonlinear Schr\"odinger (NLS) equation, is transformed by NFT to a linear memoryless channel. As a result, nonlinear frequency division multiplexing (NFDM) was proposed \cite{yousefi1,yousefi2,yousefi3}, where, rather than on the time domain, the data is mapped on either of the two available nonlinear spectra defined by NFT, namely, continuous spectrum (CS) \cite{prilepsky2014,yousefi2016nonlinear,iman} and discrete spectrum (DS) \cite{hari2016,haribi,buelow7,bulow2015experimental,dong2015nonlinear,wai,shevchenko2016lower,alex}. Simultaneous signaling on both CS and DS has been also recently investigated \cite{iman,aref2016demonstration}. Due to the complex nature of the problem, still many aspects of NFDM systems are unknown and are being investigated. For instance, recently, efficient numerical methods have been proposed for fast forward and backward NFT \cite{wahls2015fast,wahls2016fast,civelli2015numerical}, and periodic NFT was proposed in \cite{kamalian2016periodic1,kamalian2016periodic2}. 

Mapping the data on CS is an approach very similar to orthogonal frequency division multiplexing (OFDM), but with nonlinear Fourier transform instead of ordinary linear Fourier transform. Despite the interesting linearizing feature of NFT in ideal conditions, modeling the system in the presence of perturbation is not trivial. In \cite{nature,imanJLT,iman}, it was demonstrated that the CS channel is almost memoryless, yet the noise in nonlinear spectral domain was observed to be non-Gaussian and signal-dependent. Nevertheless, remarkable capacity for CS channel was predicted \cite{nature,imanJLT}, which promises the possibility of achieving the ultimate capacity of optical fiber if all degrees of freedom (i.e., CS and DS) are exploited efficiently.  

Different techniques were used for mapping data symbols on CS and detecting the received symbols. In \cite{iman,imanJLT,yousefi2016nonlinear,yangzhang2016nonlinear}, data is directly mapped on the CS and transformed to the time domain using inverse nonlinear Fourier transform (INFT) before transmission over the fiber. At the receiver side, NFT is applied to transfer back to the signal space in CS, where detection is performed. In a series of other research papers \cite{prilepsky2014,le2014nonlinear,le2015nonlinear,le2016demonstration,le2016modified}, nonlinear inverse synthesis was proposed and investigated numerically and experimentally. In this method, the linear spectrum of a predefined time domain signal (e.g., OFDM or Nyquist) rather than the signal itself is mapped on CS. The important point is that each of the aforementioned methods manifests different behavior and performance level, but this has not been investigated thoroughly in the literature. 

An important characteristic of the CS channel in NFDM systems, similar to the traditional optical fiber communication system, is the data rate limitation due to chromatic dispersion. In the absence of DS, there is no mechanism, which can limit the induced dispersion before performing NFT at the receiver. Thus, usually large temporal broadening is observed in NFT systems based on CS signaling. Since interference of two neighboring data packets in time domain cannot be removed after NFT operation, guard intervals between data packets in time domain are required to avoid error due to the interference. For instance, a guard time of 17 ns, accounting for dispersion induced memory plus a 20\% margin, was used for transmitting a data packet of 2 ns long in \cite{le2015nonlinear}. In some studies\cite{le2014nonlinear,le2015nonlinear,yangzhang2016nonlinear}, the effect of such guard intervals was not taken into account, when the achievable data rate was calculated. However, this assumption only gives a reasonable estimate if very long data packets are considered leading to a substantial increases in the complexity and the delay at the receiver. In \cite{imanCLEO}, we proposed a simple precompensation method that can decrease the induced dispersion by 50\%. Nevertheless, a shorter guard interval is still required and its effect on the design of the NFDM system needs to be studied.  

In this paper, first, the effect of dispersion on NFDM systems is investigated. It is demonstrated that better performance and higher bit rate can be achieved by properly choosing the system parameters and minimizing the effect of dispersion. Then, by considering the specific characteristics of the noise in nonlinear spectral domain, we propose three different signaling methods for performance improvement and compare them with the conventional method in \cite{yousefi3} in terms of error performance, distance reach, and complexity. Taking into account the dispersion effect, we can investigate data rates in bits per second rather than bits per symbol. In particular, it is demonstrated that 9.6 Gbps can be transmitted over 7100 km using 26 GHz bandwidth.

\section{Preliminaries}
In this section, the channel modeling and theoretical principles of NFDM systems are briefly discussed \cite{iman,imanJLT,yousefi1,yousefi2,yousefi3,prilepsky2014,ablowitz}. 
\subsection{Optical fiber communication}
Propagation of the complex envelope of a narrowband optical field in a standard single-mode fiber can be described by the stochastic nonlinear Schr{\"o}dinger (NLS) equation\cite{agrawal0}. Assuming the fiber loss to be perfectly compensated by ideal distributed amplification, the NLS equation can be described as
\begin{equation} \label{NLS}
\frac{\partial Q(T,l)}{\partial l}= -\frac{j\beta_2}{2}\frac{\partial^2Q(T,l)}{\partial T^2}+j\gamma Q(T,l)\left|Q(T,l)\right|^2+N(T,l),\;\;0\leq l\leq L,
\end{equation} 
where $Q(T,l)$, $l$, and $T$ respectively represent the complex envelope of the optical field, distance in $\mathrm{km}$, and time in seconds ($j=\sqrt{-1}$). $N(T,l)$ represents the amplified spontaneous emission (ASE) noise added by the amplifiers, which is a white Gaussian process with autocorrelation
\begin{equation}
\mathsf{E}\{N(T,l)N^*(T',l')\}=\frac{N_{ASE}}{L}\; \delta(T-T')\delta(l-l'),
\end{equation}
where $\delta(.)$ is the Dirac delta function, and $N_{ASE}=\alpha Lh\nu_sK_T$ is the accumulated spectral density of noise at fiber length $L$. Here, $\beta_2$, $\gamma$, $\alpha$, $h\nu_s$, and $K_T$ are respectively chromatic dispersion coefficient, nonlinearity parameter, fiber loss, average photon energy at carrier frequency $\nu_s$, and phonon occupancy factor. Throughout this paper, we consider the focusing case $\beta_2<0$ without any kind of dispersion compensation. It is also assumed that the effect of higher order dispersions can be ignored. The NLS equation can be normalized in the form of 
\begin{equation} \label{nls2}
j\frac{\partial q(t,z)}{\partial z}=\frac{\partial^2q(t,z)}{\partial t^2} + 2 |q(t,z)|^2q(t,z)+n(t,z),
\end{equation}
by normalization rules 
\begin{equation} \label{norm}
q=\sqrt{\gamma L_D}Q,\;\;\;z=\frac{l}{2L_D},\;\;\;t=\frac{T}{T_0},
\end{equation}
where the normalizing parameter $T_0$ can be chosen independent of other parameters. Also, we have $L_D=T_0^2/|\beta_2|$. Note that the variables in the normalized NLS equation (\ref{nls2}) are unitless.
\subsection{Nonlinear Fourier transform}
Using inverse scattering method \cite{ablowitz}, NFT is defined, which transforms the time domain optical signal into scattering data $\{\rho(\lambda,z), \{C_m(z)\}_{m=1}^M,\{\lambda_m\}_{m=1}^M\}$, which evolve linearly along the fiber in nonlinear spectral domain. The parameter $M$ represents the number of eigenvalues. The continuous spectrum (CS) $\rho(\lambda,z)$ is defined on the real axis $\lambda \in \mathbb{R}$, and the discrete spectrum $C_m(z)$ is defined on the upper half complex plane $\lambda_m \in \mathbb{C^+}$. It can be shown that, in a noise-free scenario, the eigenvalues are preserved along the fiber, and the evolution equations for continuous and discrete spectra are expressed as \cite{ablowitz}
	\begin{equation}\label{evol1}
	\lambda_m(z)=\lambda_m(0)=\lambda_m,\;\;\; m=1, \cdots, M,
	\end{equation}
	\begin{equation}\label{evol2}
	C_m(z)=C_m(0)e^{-4j\lambda_m ^2z},\;\;\; m=1, \cdots, M,
	\end{equation}
	\begin{equation}\label{evol3}
	\rho(\lambda ,z)=\rho(\lambda ,0)e^{-4j\lambda ^2z}.
	\end{equation}
It can be seen from (\ref{evol3}) that for CS the evolution equation is only a phase shift since $\lambda \in \mathbb{R}$. The (forward) NFT is performed by solving the so-called Zakharov-Shabat eigenvalue problem, and inverse nonlinear Fourier transform (INFT) can be performed by solving Gelfand-Levitan-Marchenko (GLM) integral equations \cite{ablowitz}. A number of other methods have been also proposed for INFT \cite{yousefi2016nonlinear,wahls2016fast}, and there exist several numerical methods in the literature for NFT and INFT operations \cite{yousefi2,iman,haribi,yousefi2016nonlinear,wahls2016fast}. 

\begin{figure}[h]
	\centering
	\includegraphics[width=0.8\textwidth]{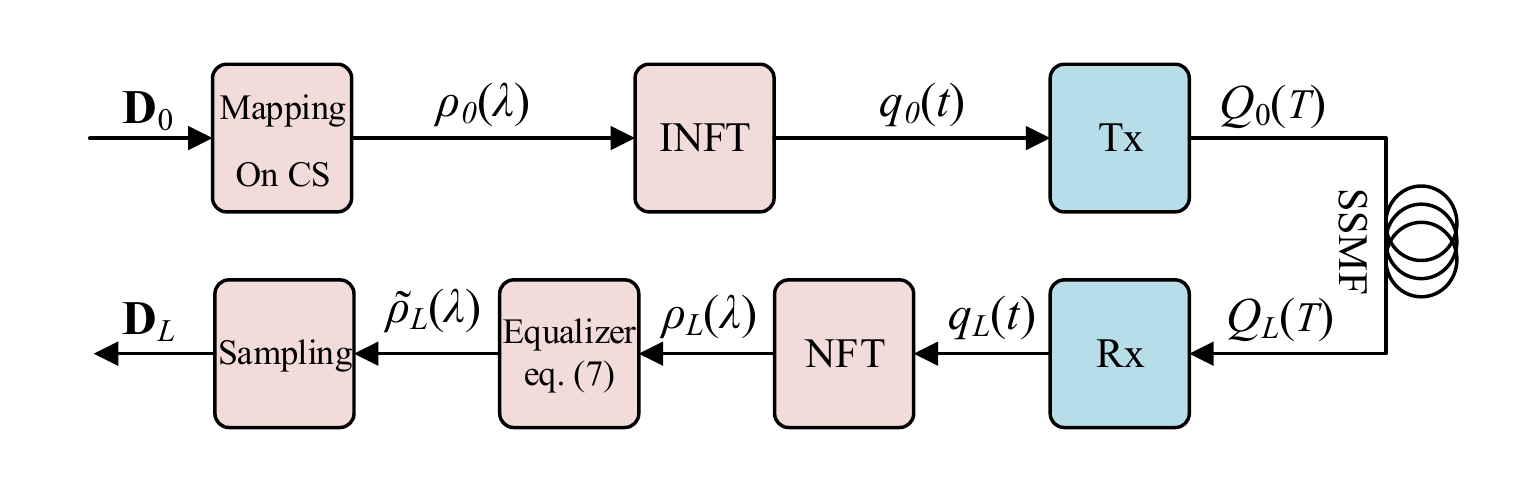}
	\caption{\small Block Diagram of a NFDM system with direct mapping on CS.}
	\label{blockdiagram}
\end{figure}

The block diagram of a typical NFDM system with direct mapping of data on CS is demonstrated in Fig.\ref{blockdiagram}. In this paper, we consider only CS, which correspond to the dispersive radiation waves. In other words, $\{C_m(z)\}_{m=1}^M$ and $\{\lambda_m\}_{m=1}^M$, which correspond to solitonic waves, are zero throughout this paper. In spite of the fact that mapping data only on CS is similar to OFDM, still many aspects of it are  unknown and need to be investigated. For simplicity in notation, the dependence to distance is dropped and shown as subscript. For instance, $q_L(t)=q(t,L)$ for fiber length $L$. In Fig. \ref{blockdiagram}, $\mathbf{D}_0$ is the vector of $K$ symbols, which are driven from the binary data using a specific modulation format. The symbols in $\mathbf{D}_0$ are then converted to an oversampled continuous waveform $\rho_0(\lambda)$ using a pulse shaping filter (e.g., raised-cosine or sinc pulses). Note that since the integral equations of INFT are solved digitally in practice, oversampling is essential for minimizing numerical errors. Then, INFT is applied to generate the time domain signal $q_0(t)$. The optical transmitter (Tx) and receiver (Rx) consist of all operations needed for launching the signal into the fiber and detecting it as a digital signal, such as (de)normalization in (\ref{norm}), optical filtering, and analog/digital conversion. Standard single mode fiber (SSMF) and ideal distributed amplification are assumed. At the receiver, after NFT operation and removing the phase shift introduced according to (\ref{evol3}), the noisy CS signal $\tilde{\rho}_L(\lambda)$ is obtained, from which the data symbols can be recovered. Throughout the paper, we refer to the conventional signaling method depicted in Fig. \ref{blockdiagram} \cite{yousefi1,yousefi2,yousefi3,yousefi2016nonlinear,yangzhang2016nonlinear,imanJLT,iman} as "direct mapping on CS" and use it as a benchmark for performance comparisons among proposed signaling techniques. 
 
We use the Ablowitz method for NFT, and discretization of Marchenko equations and using the symmetry of the resulted matrix to implement INFT operation \cite{iman}. Moreover, the split-step Fourier method is applied for simulating the propagation of signal along the fiber. Throughout this paper, $T_0=0.1$ ns is chosen. More details about our simulation method can be found in \cite{imanJLT}.

\subsection{Propagation of the continuous spectrum in the presence of noise} \label{sec_model}
In the low noise scenario, perturbation theory can be used to determine the characteristics of noisy CS \cite{nature,imanJLT}. The channel model in the nonlinear spectral domain can be expressed as \cite{yousefi3}
\begin{equation}\label{model}
	\tilde{\rho}_L(\lambda)=\rho_0(\lambda)+\eta_L(\lambda),
\end{equation}
where $\tilde{\rho}_L(\lambda)=\rho_L(\lambda)\exp(4j\lambda ^2z)$ refers to noisy CS signal after phase shift removal according to (\ref{evol3}), and $\eta_L(\lambda)$ is a complex generally non-Gaussian noise, the variance of which is dependent on the initial signal as $\mathsf{E}\{|\eta_L(\lambda)|^2\}=f[|\rho_0(\lambda)|]$. The signal-dependency of the noise in CS was studied by asymptotic analysis and numerical simulation in \cite{imanJLT}, and by applying perturbation theory in \cite{nature}. It is shown that $f[|\rho_0(\lambda)|]$ can be approximated by a polynomial of order 4 \cite{imanJLT,nature}. This means that signals with higher amplitudes experience higher noise variance. Furthermore, the real and imaginary parts of the noise in CS may not be assumed independent after propagation over the nonlinear optical fiber unlike the conventional linear channels, and they can be modeled \cite{imanJLT} as $\mathsf{E}\{\Re [\eta_L(\lambda)]^2\}=\mathsf{E}\{\Im[\eta_L(\lambda)]^2\}=f [|\rho_0(\lambda)|]/2$ . However, in many practical scenarios, the noise can be assumed white, i.e., $\mathsf{E}\{\eta_L(\lambda)\eta_L^*(\lambda')\}=f[|\rho_0(\lambda)|]\delta(\lambda-\lambda')$. The linearizing effect of NFT makes the channel model of a nonlinear optical fiber much less complex in CS as in  (\ref{model}) than in time domain as in (\ref{NLS}). Nonetheless, due to the special statistical behavior of the noise, novel techniques are required for capacity estimation as we addressed in \cite{imanJLT} and signaling, which is the subject of this paper.

In this paper, we use complex modulation using sinc pulses as expressed in (\ref{sinc}), where $D_0^i$ is the $i$th symbol from vector $\mathbf{D}_0$.
 
\begin{equation} \label{sinc}
\rho_0(\lambda)= \sum_{i=1}^{K} D_0^i \mathrm{sinc} (\frac{K}{\Lambda}\lambda+\frac{K}{2}-\frac{2i-1}{2}).
\end{equation}
The width of CS signal in nonlinear spectral domain is called nonlinear spectral width and is denoted by $\Lambda$. Note that the NFT operation at the receiver is only required at $K$ symbol points $\lambda_i=-\frac{\Lambda}{2}+(2i-1)\frac{\Lambda}{2K}$. 

For a given input symbol vector $\mathbf{D}_0$, the question arises that in what rate should the symbols be placed in nonlinear spectral domain. In other words, how should we choose the nonlinear spectral width of the signal, $\Lambda$, and if there is any optimal value for it? Note that this question originally arises since we are signaling in the nonlinear spectral domain, which does not have a linear duality with the time domain unlike conventional OFDM systems. To answer this question, the effects of chromatic dispersion on data rate and performance of NFDM systems are investigated in section \ref{sec_disp}.  Furthermore, due to the signal-dependency of the noise in this domain, conventional signal processing techniques are no longer as efficient, and thus, new techniques are studied in section \ref{sec_sig}.

\section{Dispersion effects and the optimum value of nonlinear spectral width}\label{sec_disp}
By mapping the data only on CS, which is the focus of the current paper, soliton formation is suppressed. In the absence of DS (e.g. solitons), the mechanism of cancellation of dispersion with nonlinearity no longer exists, and consequently excessive temporal broadening is expected. Therefore, large guard bands are necessary to avoid mixing of neighboring signals \cite{le2014nonlinear,le2015nonlinear,le2016demonstration}. This reduces the effective data transmission rate in bits per second. In this section, the effects of chromatic dispersion on NFDM communication systems are evaluated. Note that, throughout this paper, "bandwidth" refers to the actual linear bandwidth of time domain signal $Q_l(T)$ rather than the width of signal in the nonlinear spectrum CS. Furthermore, the temporal width and bandwidth are calculated at the window, which consists 99.5\% of signal energy. Using this definition, the numerical error due to truncation of the time domain signal is negligible. 

Chromatic dispersion is an important effect, which cannot be avoided in time domain in current NFDM systems because, based on the inverse scattering theorem \cite{ablowitz}, the signal in time domain needs to be separated from consecutive signals so that the NFT operation works properly and linearizes the channel. The effect of dispersion is removed only after NFT operation (i.e., phase shift in the CS as in (\ref{evol3})), and thus the broadening in time domain cannot be ignored. Therefore, it is beneficial to know how the broadening can be minimized in order to maximize the effective data rate. As an initial estimate on the dispersion effect, we consider the linear fiber channel as in \cite[Chapter 3.4]{agrawal0}, where broadening factor is derived for Gaussian pulses assuming a linear fiber. It is shown that an optimum initial temporal pulse width (or pulse rate) exists that minimizes the broadening effect of chromatic dispersion while the dispersion effect increases for a higher bandwidth and fiber length. The same phenomena can be also observed for the NFDM system when the data is modulated on the CS. 

For a given input symbol vector $\mathbf{D}_0$ with fixed number of symbols $K$, the signal bandwidth is directly related to the nonlinear spectral width $\Lambda$ \cite{imanJLT}. Similar to the argument about a single Gaussian pulse in the linear regime, it is expected that, for a fixed $K$, $\Lambda$ (or, in other words, the symbol rate) can be adjusted to minimize the received temporal width of signal. The existence of an optimum nonlinear spectral width $\Lambda$ in NFDM systems can be shown through the asymptotic analysis presented in \cite{imanJLT}. Assuming the data is modulated on CS with the nonlinear spectral width $\Lambda$, the initial temporal width of the signal in time domain (after INFT and at $L=0$) can be estimated as $\widetilde{\Delta T_0} =T_0\pi K / \Lambda$ for low-amplitude signals by approximating INFT as a linear Fourier transform. When the amplitude of the signal increases this temporal width increases and cannot be accurately determined using the linear estimation any more. Defining the parameter $r$ as the ratio between the actual temporal width after INFT compared with the linear estimation above (i.e., $r={\Delta T_0}/\widetilde{\Delta T_0} $), the actual temporal width can be expressed as $\Delta T_0 =rT_0\pi K/ \Lambda $, where $r$ has been shown to be increasing as a function of the signal power in nonlinear spectral domain and independent of $\Lambda$ \cite{imanJLT}. In asymptoticly long fibers, the received temporal width at length $L$ denoted by $\Delta T_L$ can be expressed by \cite{imanJLT}
\begin{equation} \label{asymp_len}
\Delta T_L=\frac{rT_0 \pi K }{\Lambda}+4zT_0 \Lambda,
\end{equation}
where $z$ is defined in (\ref{norm}). It can be readily seen from (\ref{asymp_len}) that the received temporal width $\Delta T_L$ can be minimized at $\Lambda=\sqrt{r \pi K/ 4 z}$. Consequently, for a given sequence of $K$ CS symbols containing fixed amount of information, an optimal $\Lambda$ (or equivalently symbol rate) can be selected to minimize the required guard interval  leading to higher packet rates and effectively higher data rates in bits per second.

\begin{figure}[h]
	\centering
	\includegraphics[width=0.7\textwidth]{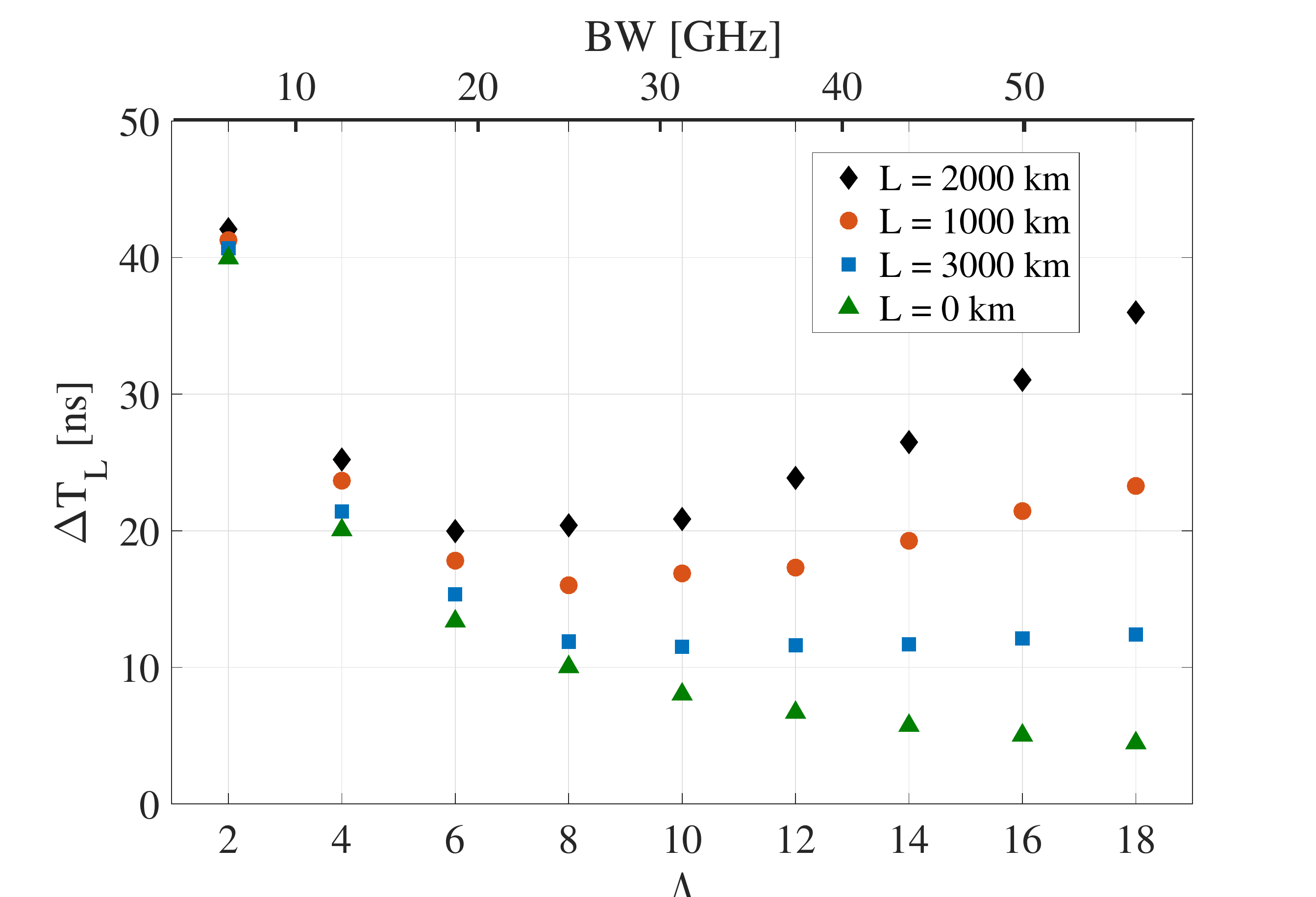}
	\caption{\small The temporal width of the propagated signal after distance $L$ for different BW and $\Lambda$.}
	\label{TL}
\end{figure}

It can be also confirmed numerically that an optimal value exists for spectral width $\Lambda$ (or bandwidth BW) that minimizes $\Delta T_L$ in practical (non-asymptotic) scenarios. Assuming that $K=128$ random QPSK modulated symbols are mapped on CS using sinc pulses as in (\ref{sinc}), Fig. \ref{TL} shows the existence of the optimal bandwidth for different fiber lengths. The optimal bandwidth decreases for longer fibers. Furthermore, it is confirmed by simulation that the initial width (i.e., $\Delta T_L$ at $L= 0 $) decreases when the bandwidth increases, and $r$ is constant for all values of spectral width $\Lambda$. Signal power in nonlinear spectral domain used in this simulation corresponds to $r=1.99$. It is worth mentioning that, for instance, the optimal value of $\Lambda$ for $L=2000$ km and $r=1.99$, calculated based on the asymptotic formula (\ref{asymp_len}), is obtained as 9.76, which is close to the actual value obtained by simulation in Fig. \ref{TL}. 

It should be noted that, as stated in \cite[Table IV]{imanJLT}, by increasing $K$, the effect of dispersion can be reduced and higher overall data rates are achieved even when $\Lambda$ remains fixed rather than being adjusted to the optimal value for the new $K$. On the other hand, the computational complexity increases for larger $K$, and thus $K$ and $\Lambda$ (or bandwidth) should be optimized based on the trade-offs between performance metrics such as computational complexity and effective data rate. Such optimization problems will be the subject of future research.

Another crucial consequence of minimizing the received temporal width by choosing the optimal value for $\Lambda$ is the noise reduction in nonlinear spectral domain, which may lead to higher achievable data rates by reducing the number of errors. In NFDM systems, each single sample of data at a nonlinear frequency $\lambda$ receives signal and noise contributions from all the components within the received temporal width of signal. Therefore, as stated in \cite{nature}, a larger received temporal width results in higher noise in nonlinear spectral domain. This is similar to ordinary linear communication systems, in which the noise power in time domain is directly related to the signal bandwidth. In effect, if the received temporal width, over which the NFT operation is performed, shrinks, the effective noise added to CS in nonlinear spectral domain decreases.

\begin{figure}[h]
	\centering
	\includegraphics[width=0.7\textwidth]{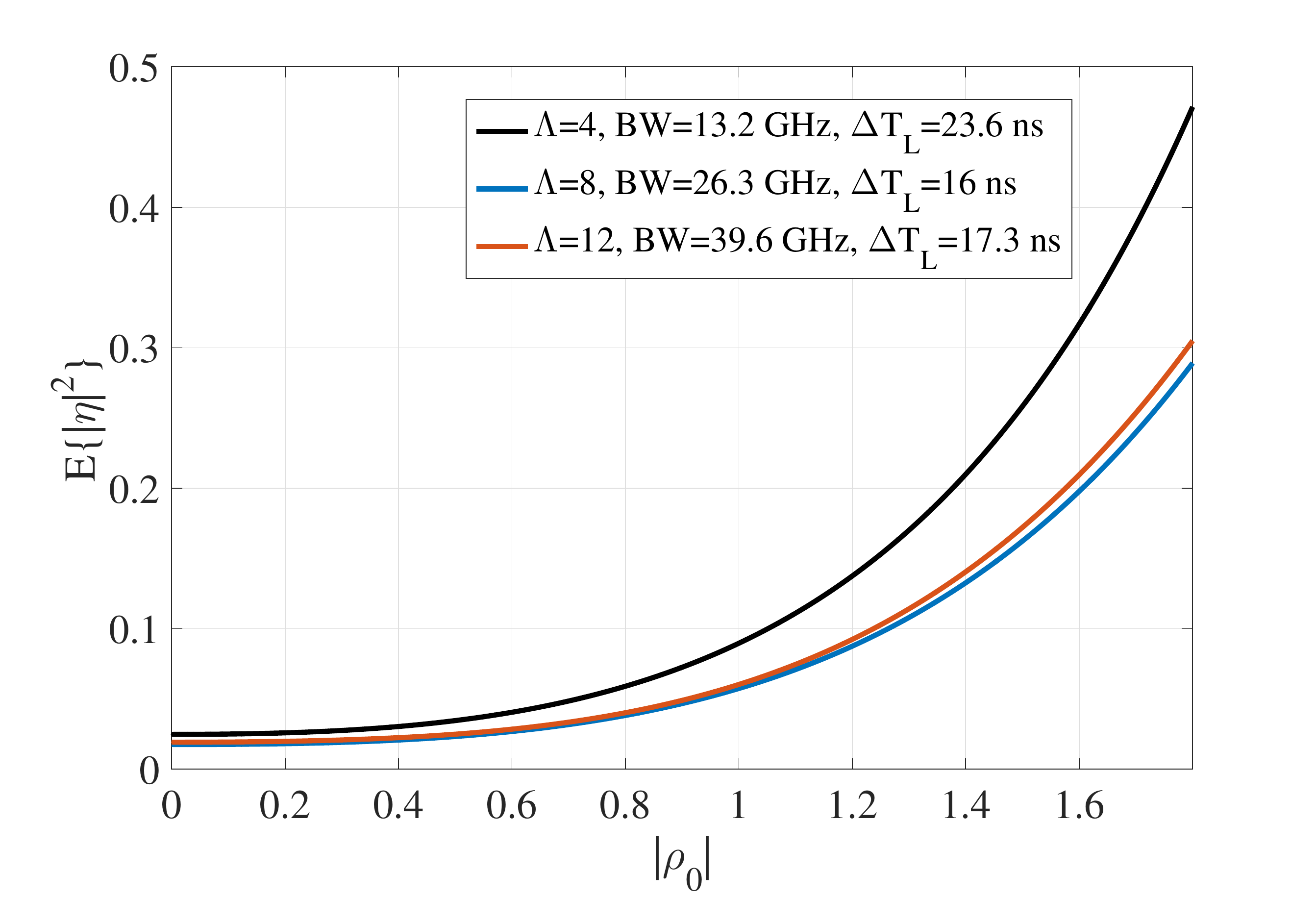}
	\caption{\small The variance of the noise in nonlinear spectral domain for different $\Lambda$}
	\label{noise2}
\end{figure}

Here, a simulation is performed for different values of $\Lambda= 4, 8, 12$, which shows that the largest received temporal width (corresponding to the smallest $\Lambda= 4$) results in higher noise in nonlinear spectral domain. Figure \ref{noise2} illustrates this result by plotting the noise variance for different values of $\Lambda$ against the amplitude of the signal, which also shows the signal-dependency of the noise. The results in Fig. \ref{noise2} demonstrate that the noise induced in nonlinear spectral domain is directly related to the received temporal width at any amplitude of CS. This figure is obtained by calculating the noise variance for more than $2\times10^6$ symbols ($2\times10^7$ samples) in different nonlinear frequencies. Figure \ref{noise2} suggests that choosing the optimum $\Lambda$ not only minimizes the dispersion effect, but also reduces the noise. Therefore, improved error performance and achievable data rate is expected.

Further reduction in noise variance of the nonlinear spectral domain is expected if a dispersion precompensation method, similar to the technique proposed in \cite{imanCLEO}, is used. In \cite{imanCLEO}, the signal is first phase shifted by $e^{+2j\lambda ^2z}$ in CS. Then, it is transformed by INFT to time domain and transmitted through the fiber. This reduces the effect of dispersion up to two times. Consequently, the received temporal width and the required guard interval are decreased. Also, it is possible to introduce a complete precompensation by performing INFT on $\rho_0(\lambda)e^{+4j\lambda ^2z}$ at the transmitter side. It is expected that this method further reduces the noise in nonlinear spectral domain, but the required guard interval would remain equivalent to the case in which no dispersion precompensation is used. Due to the added complexity, none of these precompensation techniques are considered in the current paper. 

\section{Signaling methods}\label{sec_sig}
In this section, three different methods are introduced for performance enhancement of the continuous spectrum channel (\ref{model}) \cite{yousefi3,iman}. As a benchmark we also consider the conventional method of ``direct mapping on CS'', which refers to the block diagram shown in Fig. \ref{blockdiagram} and channel model (\ref{model}). In order to properly understand the effect of signal-dependent noise, a ring constellation is used for modulation. It should be noted that even in conventional fiber optic systems ring constellation is a popular choice \cite{capacity,sorokina2016ripple,skidin2016mitigation}.  Unless otherwise stated, $K=128$ random symbols are mapped on CS using sinc pulses as in (\ref{sinc}) for $\Lambda=8$. The fiber length is $L=2000$ km and the signal bandwidth is 26 GHz.

\subsection{Nonuniform signaling }\label{sec_sig_vnt}
Using nonuniform levels for constellation diagram is an effective method for improving the performance of a channel with signal-dependent noise \cite{sorokina2016ripple,steve}. If the statistics are known, the optimum levels can be found by solving an optimization problem such as minimization of BER. For the CS channel model (\ref{model}), unfortunately, a closed-form description of the noise statistics is not known. Therefore, deriving and solving an optimization problem for this channel would be cumbersome. Here, we propose another approach based on the variance normalizing transform (VNT) as introduced in \cite{bartlett,curtiss} for optimizing the constellation levels for NFDM systems.  

The VNT was used in \cite{imanJLT} as a tool to approximate the capacity of CS channel (\ref{model}). It has also been used for efficient signaling and capacity approximation over shot-noise-limited optical channels \cite{safari, tsiatmas,imanROF}. In general, the variance normalizing transform can be applied to a random variable, where its variance varies as a function of its mean, to generate an approximately Gaussian random variable with a variance independent of its mean. Considering (\ref{model}) and removing the dependence to $\lambda$ for simplicity, we have random variable $\tilde{\rho}_L$, with mean $\rho_0$ and variance $\mathsf{E}\{|\tilde{\rho}_L-\rho_0|^2\}=f(\rho_0)$. VNT that normalizes this random variable is defined as  
\begin{equation} \label{vnt}
T(u)=\int \frac{1}{\sqrt{f(u)}} \mathrm{d}u.
\end{equation}
The normalized (i.e., transformed) random variable $Y=T(\tilde{\rho}_L)$ has then the statistics of $\sigma_Y^2 \simeq 1$ and $\mu_Y \simeq T(\rho_0)$ for sufficiently large values of $\rho_0$. In \cite{prucnal}, it was also shown that the probability distribution of the normalized random variable tends to Gaussian distribution for a family of originally non-Gaussian probability distribution. Moreover, it was demonstrated in \cite{imanJLT} by numerical simulation that the distribution of noise on CS becomes Gaussian after applying VNT. Applying VNT as defined in (\ref{vnt}) to the noisy signal $\tilde{\rho}_L=\rho_0+\eta_L$, we will have 
\begin{equation}\label{eq_model_vnt}
Y=T(\tilde{\rho}_L)=T(\rho_0+\eta_L)\simeq T(\rho_0)+\eta_T,
\end{equation} 
where $\eta_T$ can be well approximated as a zero mean Gaussian noise with unit variance independent of the transformed signal $T(\rho_0)$. Consequently, a communication channel can be defined, in which the signal is originally generated in a \textit{transformed} domain and then mapped into the original signal domain $\rho_0$ using inverse VNT. After transmission through the channel (\ref{model}) and addition of signal-dependent noise $\eta_L$, VNT is applied, and the output signal can be expressed as $Y=T(\rho_0)+\eta_T$, which defines a conventional additive Gaussian noise channel. As in \cite[Lemma]{imanJLT}, the capacity of this transformed channel is equal to the capacity of the original channel (\ref{model}) with signal-dependent noise. Moreover, the optimal signaling techniques for the conventional additive white Gaussian noise (AWGN) channels would be also optimum for the transformed channel (\ref{eq_model_vnt}).   

The analysis above is valid for real signaling on CS and can be extended to complex signaling only if the real and imaginary channels of CS are independent. However, as mentioned in section \ref{sec_model}, when signal is modulated on CS, the real and imaginary channels are not independent \cite{imanJLT}. Moreover, although the noise on the phase of the signal is not dependent to its mean (i.e., input phase), but it depends on the input amplitude. Therefore, VNT with its form in (\ref{vnt}) can not be applied to the complex plane. However, if only real channel is used ($\Im \{\rho_0\}=0$), the noise variance would only depend on the mean (i.e., input signal), and VNT would be applicable. As in \cite{imanJLT}, the signaling can be performed in the transformed channel, but this requires the forward and inverse VNT to be performed each time a signal is transmitted and received. Alternatively, the nonuniform signal levels defined by VNT and the corresponding decision boundaries required for detection at the receiver can be determined for a known channel at the beginning of the communication and then the determined optimum levels can be used for mapping data on CS. 

\begin{figure}[h]
	\centering
	\begin{subfigure}[h]{0.35\textwidth}         
		\centering
		\includegraphics[width=\textwidth]{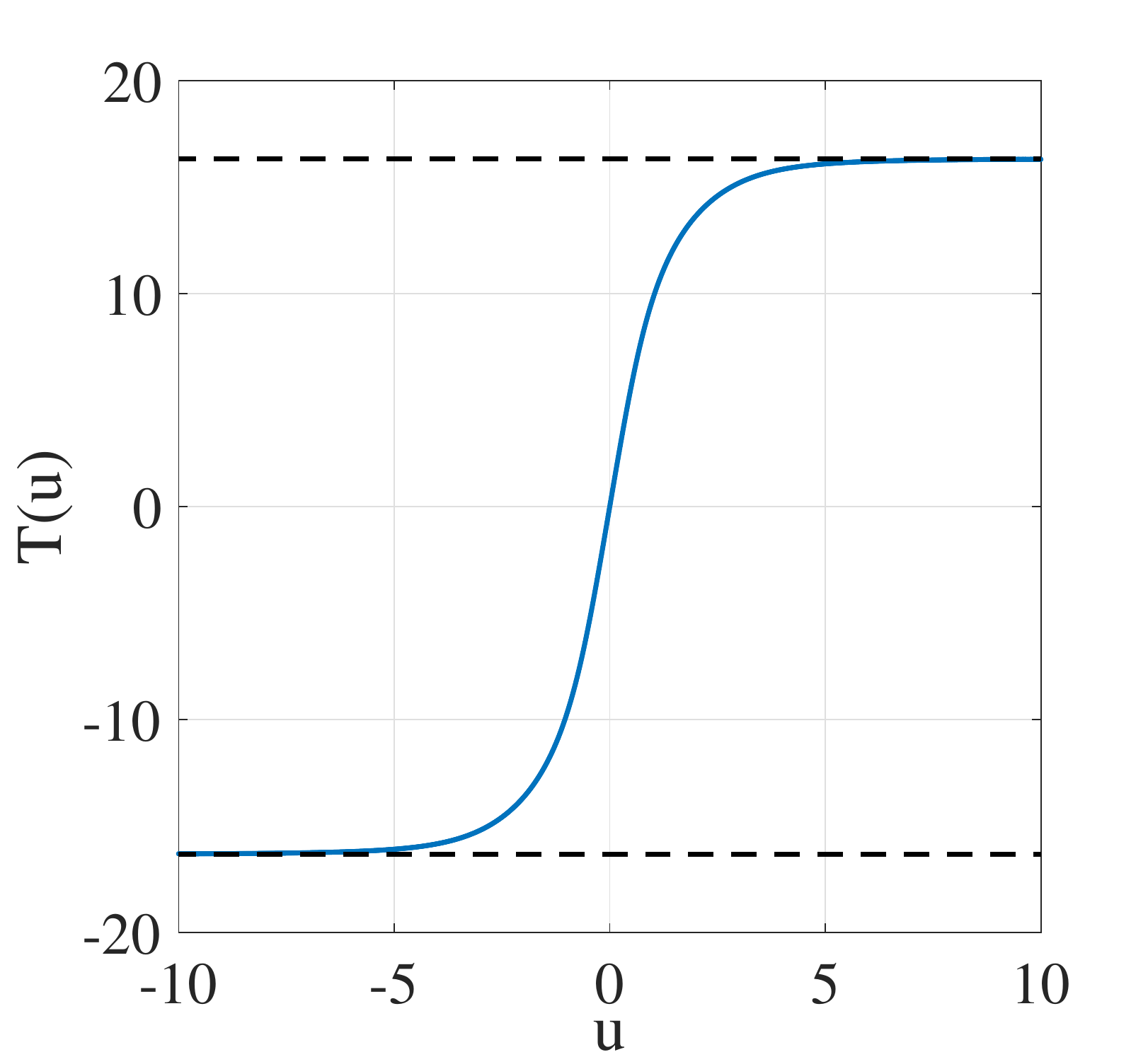}
		\caption{}
		\label{vntfig}
	\end{subfigure}%
	~ 
	\begin{subfigure}[h]{0.35\textwidth}
		\centering
		\includegraphics[width=\textwidth]{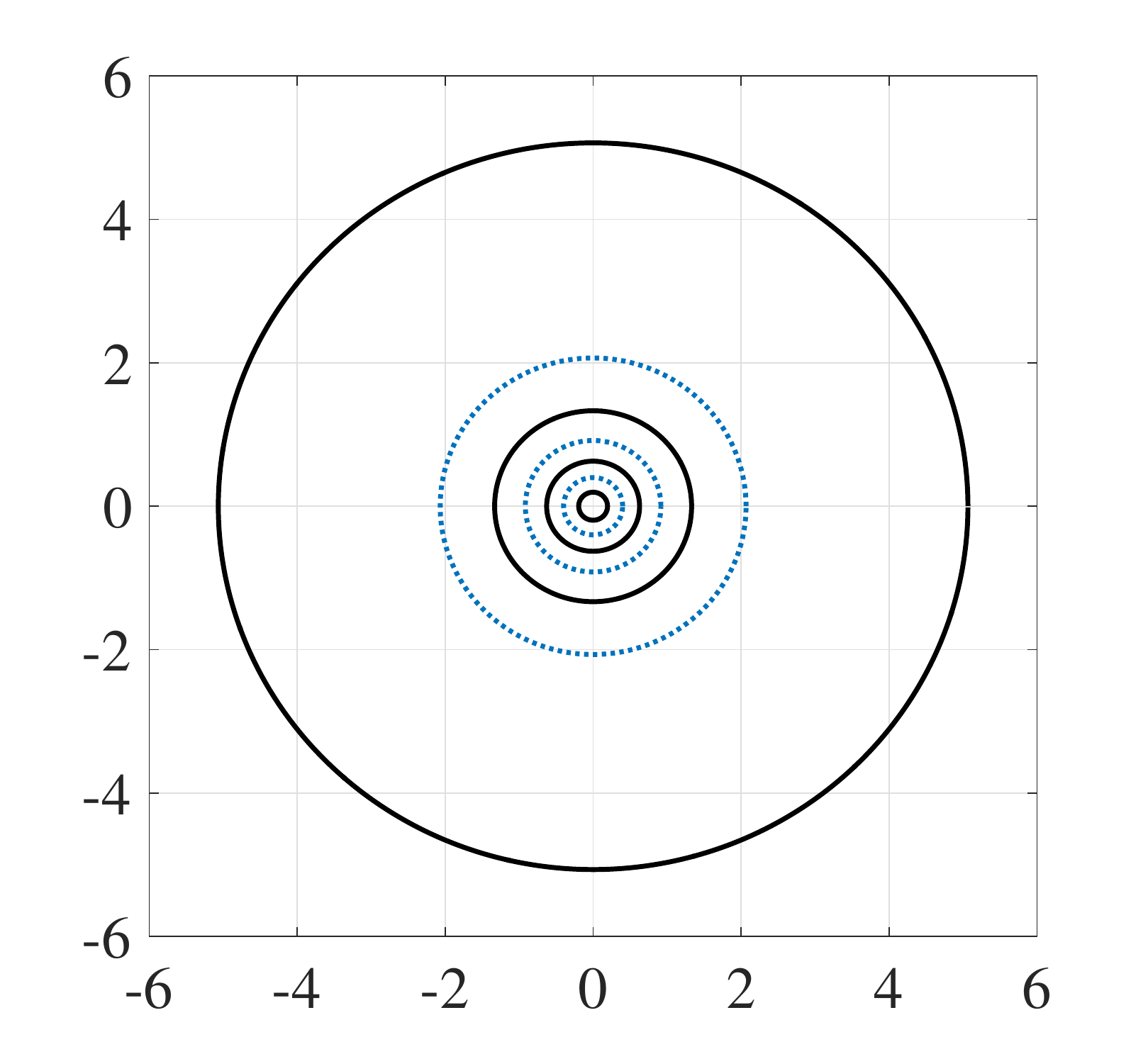} 
		\caption{}
		\label{vntring}
	\end{subfigure}%
	\caption{\small (a) Variance normalizing transform for 2000 km (b) Nonuniform levels (solid) and decision boundaries (dotted) derived from VNT.}
\end{figure}

In this paper, first it is assumed that $\Im \{\rho_0\}=0$, and uniform levels are considered for $T(\rho_0)$. Then, inverse VNT is applied to obtain the nonuniform levels for $\rho_0$. Then, the complex modulation format is defined on the rings with radius equal to the levels found. This makes our approach a sub-optimal method for complex signaling on CS because of the dependence of real and imaginary channels. Note that the number of constellation points on each ring can be chosen based on the variance of the phase of the noise and the required performance. The variance normalizing transform for the $2000$ km link is shown in Fig. \ref{vntfig}, where it is observed that $T(u)$ converges to 16.32 at $u\rightarrow\infty$ implying that a peak amplitude constraint exists for the transformed variable $T(\rho_0)$. In other words, the signal space is limited in the transformed domain because the variance of noise depends on the signal amplitude by a polynomial of order larger than 2 \cite[Theorem]{imanJLT}. For demonstration purposes, four uniform levels for $T(\rho_0)$ are chosen between 0 and 16.1, and nonuniform levels of $\rho_0$ are calculated using inverse VNT. The corresponding rings and decision boundaries are shown in Fig. \ref{vntring}. It is expected that if nonuniform levels in Fig. \ref{vntring} are used, the error rate will be lower compared to uniform levels with the same average power. This will be demonstrated in section \ref{sec_error}.

\subsection{Direct mapping on CS and filtering}\label{sec_sig_fil}
The second method, which can potentially improve the performance, is the application of linear filtering at the receiver in the original signal space, i.e., continuous spectrum. This method is similar to match filtering when detecting pulse modulation in the nonlinear spectral domain \cite{iman,yangzhang2016nonlinear} and requires oversampling of $\tilde{\rho}_L(\lambda)$ at the receiver. If INFT is applied to $\tilde{\rho}_L(\lambda)$ after removing the dispersion induced phase shift according to equation (\ref{evol3}), the signal is squeezed into the temporal width of the unpropagated signal $q_0(t)$ because of dispersion compensation in CS and while noise is still present over the uncompensated temporal width of $q_L(t)$. To get back to the signal space, we can then apply an NFT to the time domain signal after windowing the signal component distributed only within the unpropagated temporal width thereby removing the excess noise. However, this method requires complex numerical computation because of the extra NFT/INFT operations on the oversampled signal, and thus, we use ordinary Fourier transform instead of NFT, which allows for linear filtering of the CS signal. In other words, linear inverse Fast Fourier transform (IFFT) is applied to $\tilde{\rho}_L(\lambda)$, the noise out of the signal's window is removed, and then forward linear Fast Fourier transform (FFT) is applied. It should be noted that I/FFT should not be almost equivalent to I/NFT (i.e., asymptotic linear regime) to be effective in noise cancellation, as significant amount of noise is observed outside the signal interval even after IFT. In fact, employing FFT and NFT would result in different signal (and noise) power distributions. For example, as explained in section \ref{sec_disp}, so-called low-amplitude long ``tails'' \cite{prilepsky2014,wahls2016fast} typically appear after INFT while compact (finite duration) and much more uniform signal power distribution is observed after linear Fourier transform operation. Therefore, the employment of a simple constant-gain windowing operation can more effectively mitigate the noise effect for linear filtering compared to the NFT-based filtering which leads to long tails.

Assume $H(t)$ as the ideal filter with width equal to the width of $\mathrm{IFFT}\{\rho_0(\lambda)\}$, then the output can be described as
\begin{equation}\label{filter}
\hat{\rho}_L(\lambda)=\mathrm{FFT}\{H(t)\mathrm{IFFT}\{\tilde{\rho}_L(\lambda)\}\}.
\end{equation} 

\begin{figure}[h]
	\centering
	\begin{subfigure}[t]{0.25\textwidth}         
		\centering
		\includegraphics[width=\textwidth]{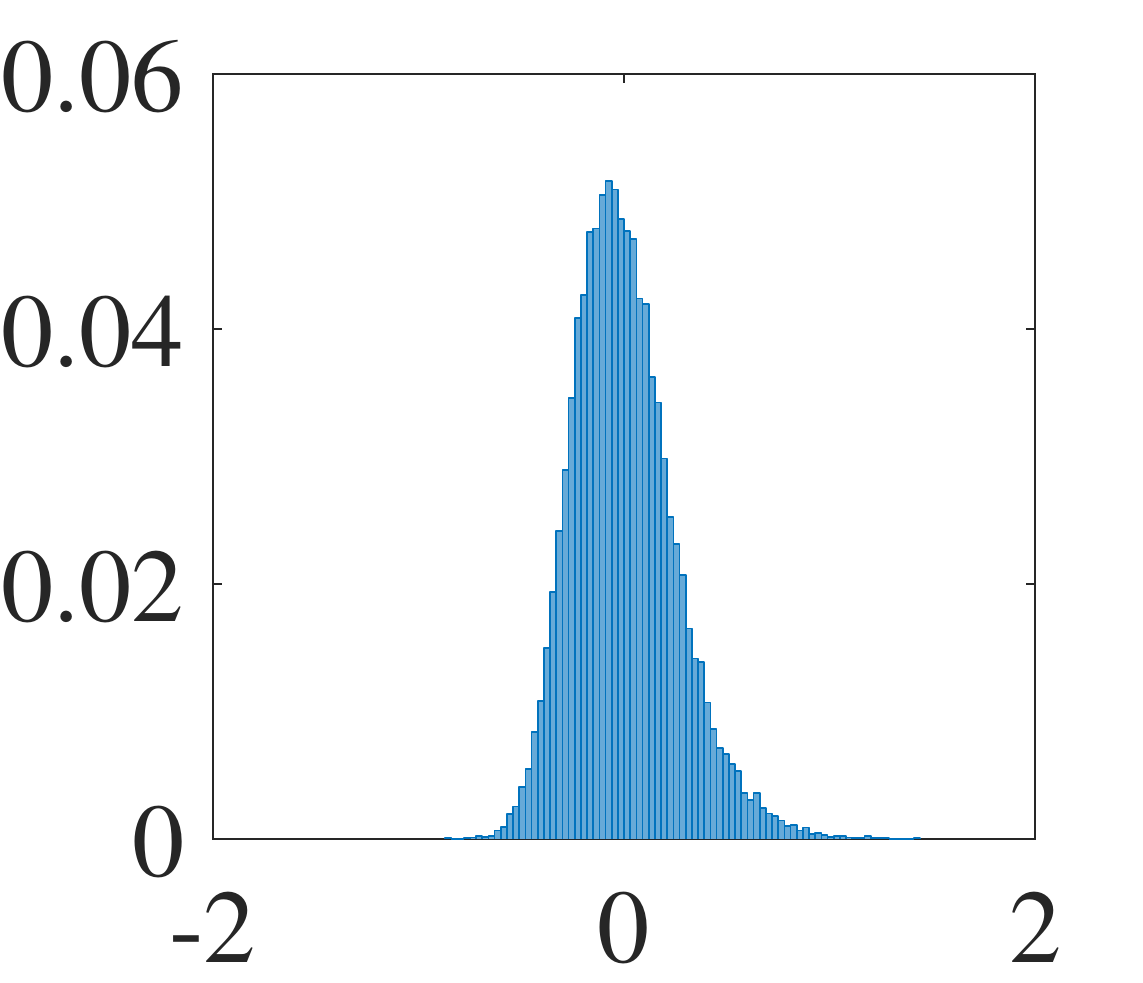}
		\caption{}
		\label{r1}
	\end{subfigure}%
	~ 
	\begin{subfigure}[t]{0.25\textwidth}
		\centering
		\includegraphics[width=\textwidth]{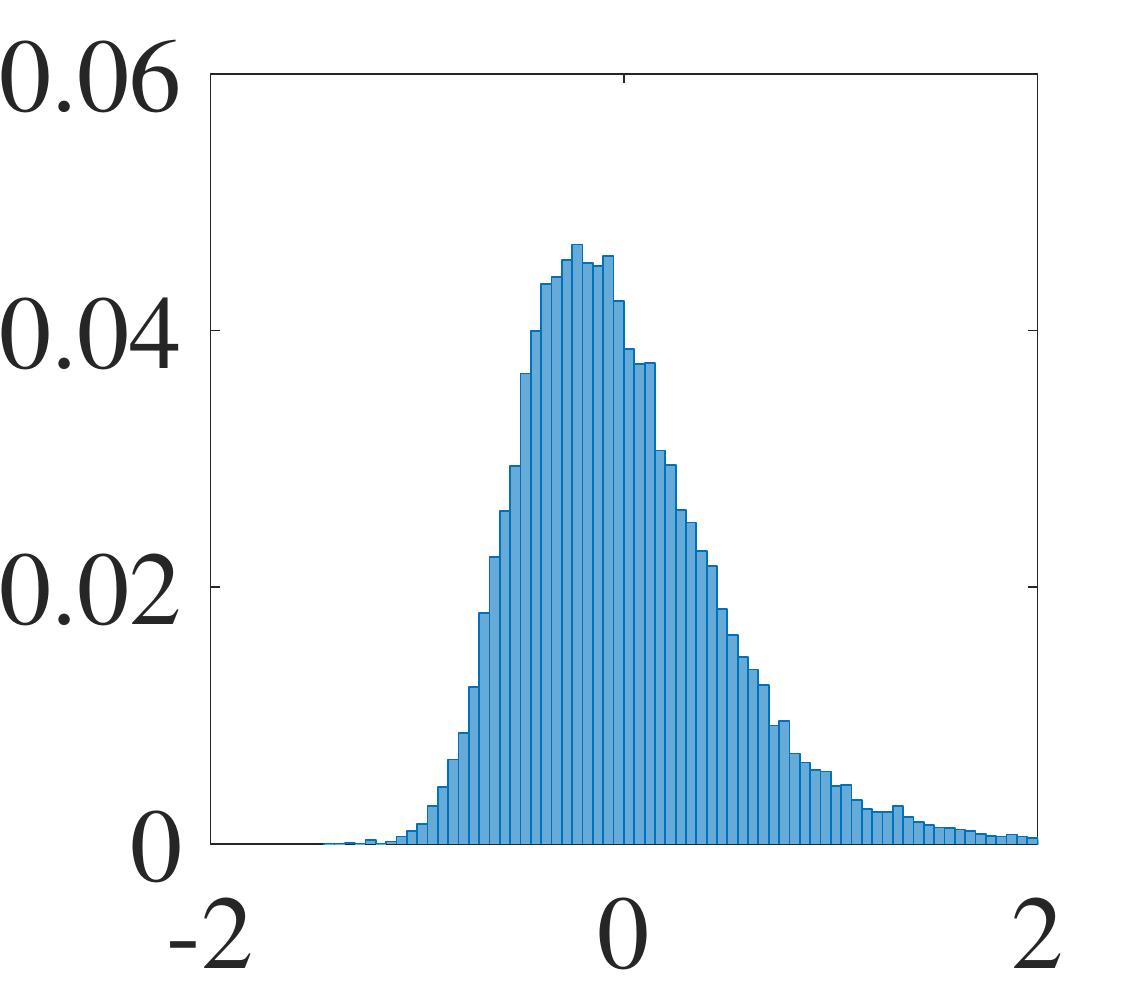} 
		\caption{}
		\label{r2}
	\end{subfigure}%
	~
	\begin{subfigure}[t]{0.25\textwidth}
		\centering
		\includegraphics[width=\textwidth]{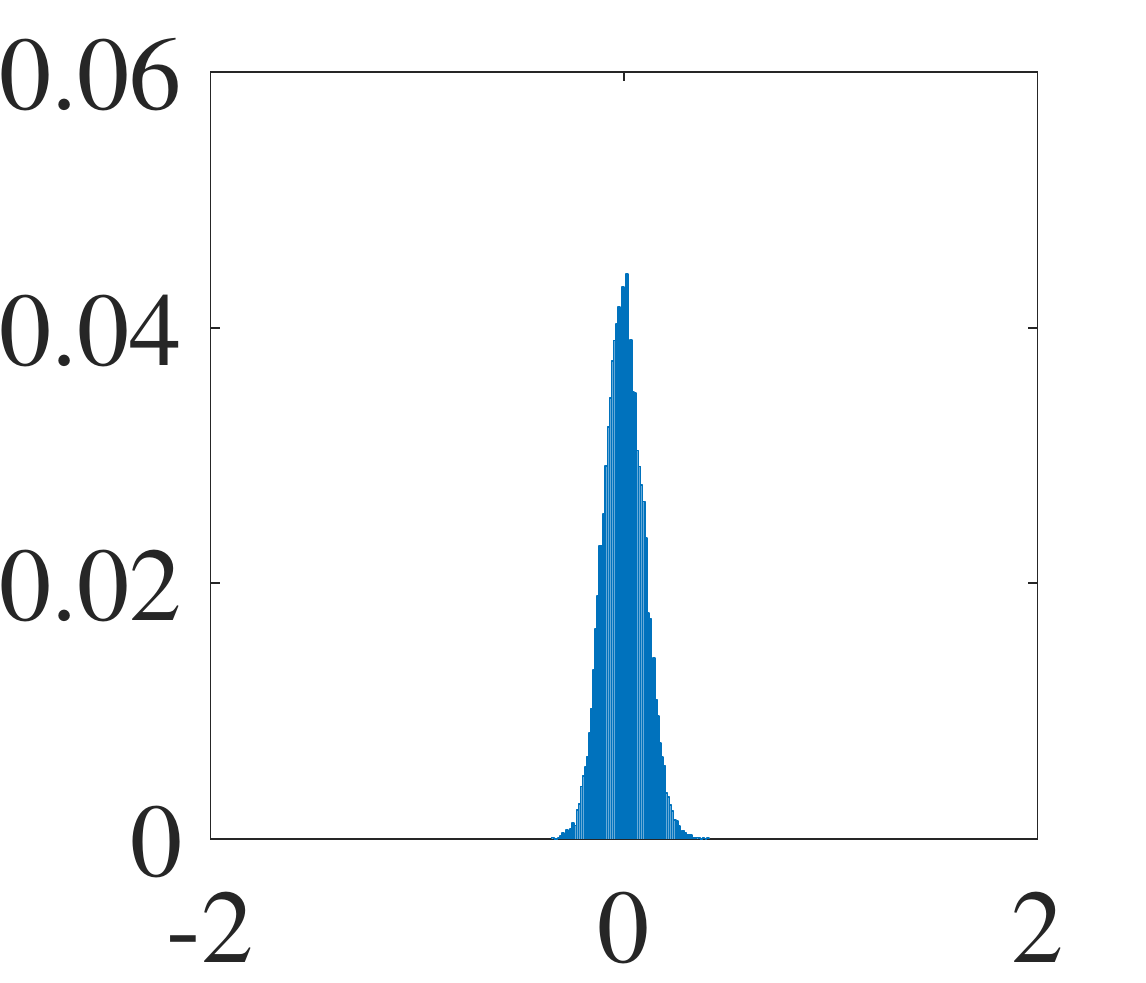} 
		\caption{}
		\label{r1f}
	\end{subfigure}%
	~ 
	\begin{subfigure}[t]{0.25\textwidth}
		\centering
		\includegraphics[width=\textwidth]{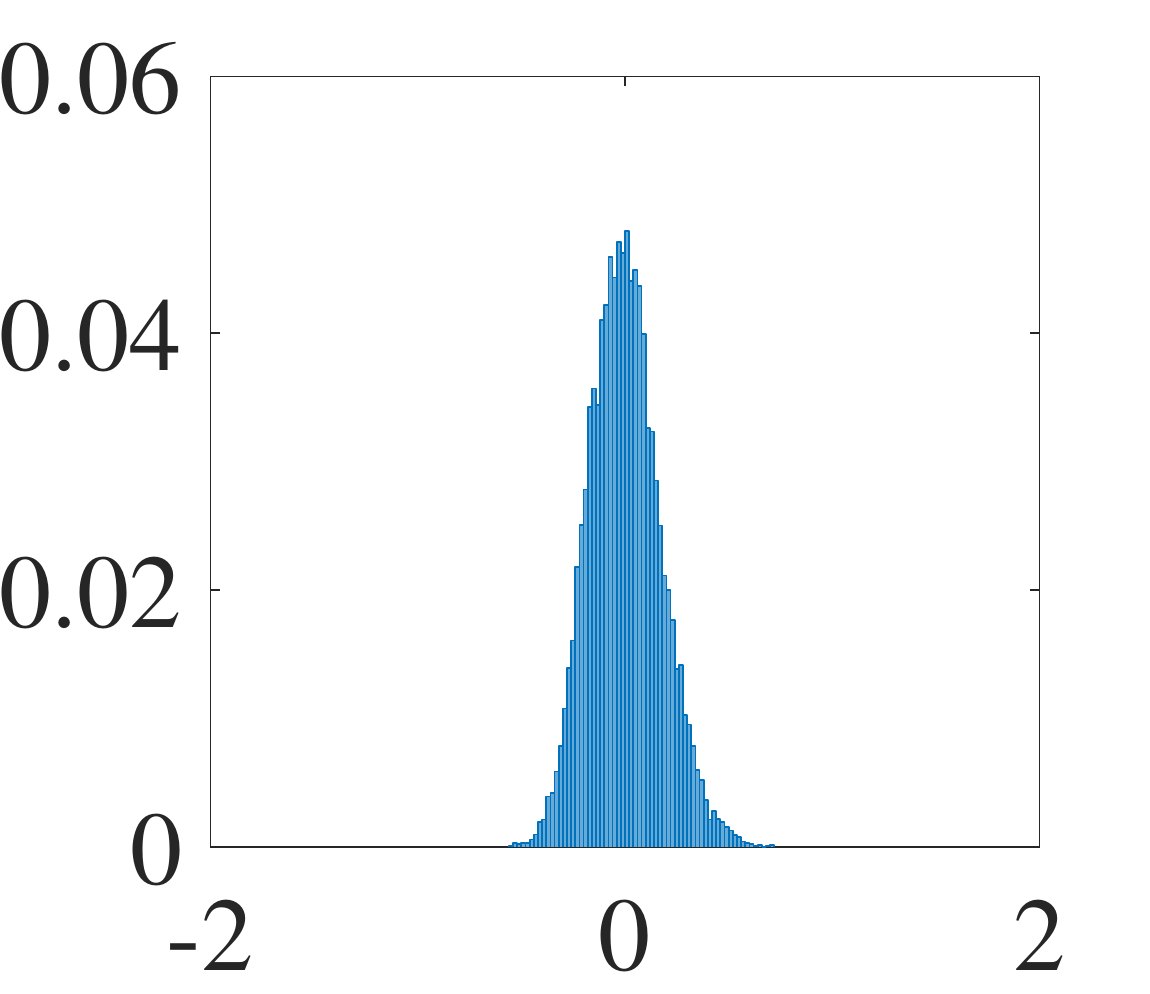} 
		\caption{}
		\label{r2f}
	\end{subfigure}%
	\caption{\small The histogram of noise on the CS signal (a)-(b) $\tilde{\rho}_L(\lambda)$ for amplitudes $\rho_0(\lambda)=1,2$, and (c)-(d) $\hat{\rho}_L(\lambda)$ (i.e., filtered signal) for amplitudes $\rho_0(\lambda)=1,2$.}
	\label{filt}	
\end{figure}

For the simulation, an oversampled version of $\tilde{\rho}(\lambda)$ by a factor of 20 is first generated using NFT, and then $\hat{\rho}_L(\lambda)$ is obtained from equation (\ref{filter}). The probability distribution of noise are demonstrated in Fig. \ref{filt} before and after filtering for signal amplitudes of  $\rho_0(\lambda)=1,2$. It can be seen that the variance of noise after filtering (Figs. \ref{r1f} and \ref{r2f}) is substantially reduced compared to the original noise (Figs. \ref{r1} and \ref{r2}). Note that although the signal-dependency of noise is not eliminated by filtering, its effect is reduced considerably. 


\subsection{GLM-based signaling}\label{sec_sig_F}
The signal modulated on the continuous spectrum $\rho(\lambda)$ is related to the time domain signal $q(t)$ by the Gelfand-Levitan-Marchenko (GLM) equation \cite{ablowitz}
\begin{equation}
K(x,y)+\int\limits_{x}^{\infty}\int\limits_{x}^{\infty} K(x,s') F^*(s+y)F(s'+s) \mathrm{d}s \mathrm{d}s'=F^*(x+y),
\end{equation}
where 
\begin{equation} \label{eq_F}
F(t)=\frac{1}{2\pi}\int_{-\infty}^{\infty} \rho(\lambda) e^{j\lambda t} \mathrm{d} \lambda
\end{equation}
is the inverse Fourier transform (IFT) of $\rho(\lambda)$. One can interpret that the GLM equation performs a nonlinear mapping on the linear inverse Fourier transform of $\rho(\lambda)$ (i.e. $F(t)$), so that the time domain signal $q(t)=-2K(t,t)$ only contains CS and no DS. This means that if GLM is applied to any arbitrary input signal, the output signal would not include any soliton components. Therefore, GLM can be used as a transformation, based on which a signaling method can be designed for the CS channel.  

\begin{figure}[h]
	\centering
	\includegraphics[width=0.8\textwidth]{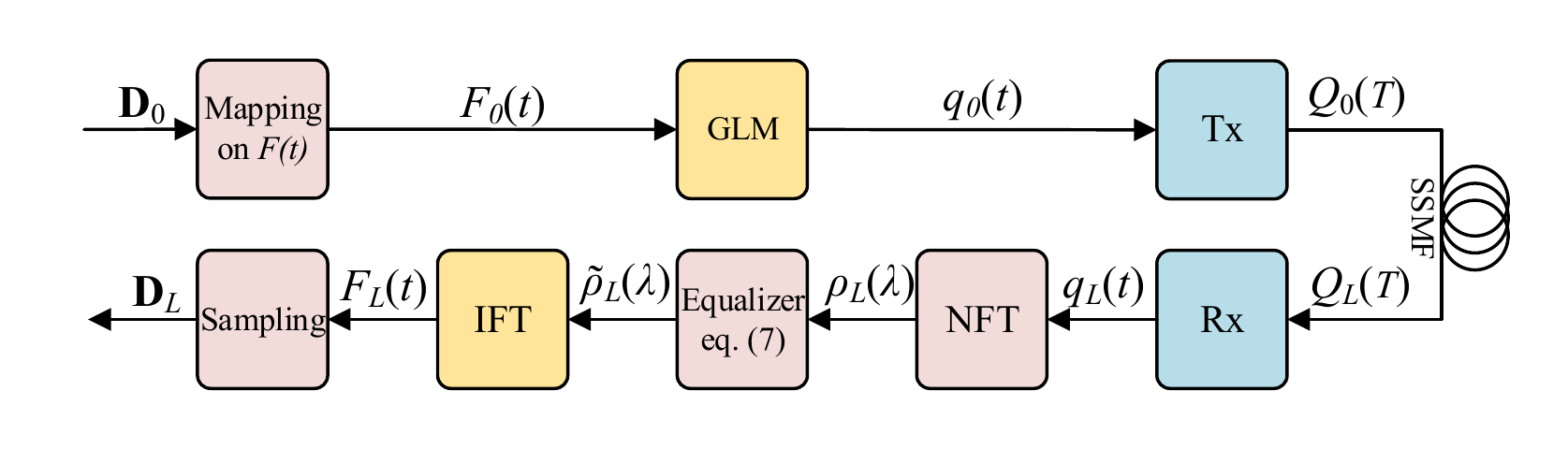} 
	\caption{\small The block diagram of the GLM-based signaling system.}
	\label{sig_F}
\end{figure}

In this section, a GLM-based signaling approach is proposed as shown in Fig. \ref{sig_F} \cite{imaneff}. The main difference of the system in Fig. \ref{sig_F} compared to the direct mapping on CS (as in Fig. \ref{blockdiagram}) is that the inverse Fourier transform operation included in INFT operation is transferred from transmitter to the receiver, and, effectively, data is mapped on $F(t)$ (i.e., linear Fourier transform of CS) defined as 
\begin{equation} \label{sincF}
F_0(t)= \sum_{i=1}^{K} D_0^i \mathrm{sinc} (\frac{K}{\tau}t+\frac{K}{2}-\frac{2i-1}{2}),
\end{equation}
where $\tau$ is the corresponding signal window in time domain determined based on system constraints, such as bandwidth and power. Consequently, an IFT block is needed at the receiver after NFT operation to return to the original signaling space, where the noisy version of (\ref{sincF}) is sampled at the center of pulses and are decoded to the original data. It should be also noted that this GLM-based signaling using sinc pulses is in fact equivalent to the nonlinear inverse synthesis method with Nyquist pulse shaping studied in \cite{le2014nonlinear,le2015nonlinear,le2016demonstration,le2016modified}. However, either in (\ref{sinc}) or (\ref{sincF}), using sinc pulse shapes are not necessary, and more efficient pulse shapes may be found by optimizing the trade-off between performance and numerical error. A similar method was also proposed independently in \cite{frumin2017new}, in which data are mapped on the \textit{kernel of GLM equations} for discrete and continuous spectrum.

Using the law of large numbers, we can show that applying (inverse) Fourier transform on the received CS diminishes the signal-dependency of the noise \cite{safari} while the probability distribution of the transformed symbols tends to Gaussian according to the central limit theorem \cite{ofdm}. Therefore, conventional modulation and signal processing techniques should perform efficiently if applied in the system illustrated in Fig. \ref{sig_F}. 

As discussed in Section \ref{sec_model}, the variance of noise sample on CS, $\eta_L(\lambda)$, depends on the input sample $\rho_0(\lambda)= \mathsf{FT}\{F_0(t)\}$. Let $\kappa$ denote the number of samples in nonlinear spectral domain, $h=\Lambda/\kappa$ be the distance between subsequent samples, and $\lambda_k$ be the $k$th sample point. The noise for GLM-based signaling is defined as $\Gamma_L(t)=F_L(t)-F_0(t)$. Therefore, for a given input signal $F_0(t)=\mathrm{IFT}\{\rho_0(\lambda)\}$ and by using the discretized version of (\ref{eq_F}), we have 
\begin{eqnarray} 
\mathsf{E} \{\Re [\Gamma_L(t)]^2 \} & \approx &\left( \frac{h}{2\pi} \right)^2 \sum\limits_{k=1}^{\kappa} \sum\limits_{p=1}^{\kappa} \mathsf{E}\{\Re [\eta_L(\lambda_k)] \Re [\eta_L(\lambda_p)]\} \cos(\lambda_k t)  \cos(\lambda_p t)\\
& = &\frac{1}{2}\left( \frac{h}{2\pi} \right)^2 \sum\limits_{k=1}^{\kappa} \sum\limits_{p=1}^{\kappa} \delta(k-p) f[|\rho_0(\lambda_k)|] \cos(\lambda_p t)  \cos(\lambda_k t)\\
& = &\frac{1}{2}\left( \frac{h}{2\pi} \right)^2 \sum\limits_{k=1}^{\kappa} f[|\rho_0(\lambda_k)|] \cos(\lambda_k t)^2 \\
& = &\frac{h\Lambda}{16\pi^2} \frac{1}{\kappa}\sum\limits_{k=1}^{\kappa} f[|\rho_0(\lambda_k)| + \frac{h\Lambda}{16\pi^2} \frac{1}{\kappa} \sum\limits_{k=1}^{\kappa} f[|\rho_0(\lambda_k)|] \cos(2\lambda_k t) \label{eq_varF0}\\
& \approx & \frac{h\Lambda}{16\pi^2} \mathsf{E} \{f[|\rho_0|]\}. \label{eq_varF}
\end{eqnarray} 

For the last equality, the law of large numbers can be used (at large $\kappa$) so that the first sum term in (\ref{eq_varF0}) can be described as the expected value of $f[|\rho_0|]$, and the second term tends to zero because of the presence of the cosine term. A similar result can be demonstrated for $\Im [\Gamma_L(t)]$, and it is observed from (\ref{eq_varF}) that the variance of $\Re [\Gamma_L(t)]$ and $\Im [\Gamma_L(t)]$ at every sampling time of $t$ are independent of signal $F_0(t)$ at that time. However, it is clear that the variance of noise $\Re [\Gamma_L(t)]$ and $\Im [\Gamma_L(t)]$ depend on average value of $f[|\rho_0|]$, indicating that higher energy (i.e., higher average power) results in larger noise variance. This will be investigated in terms of error rate in Section \ref{sec_error}. Also, according to the central limit theorem \cite{ofdm}, the probability distribution of $\Gamma_L(t)$ tends to Gaussian. This is similar to the effect observed in visible light communications where the non-Gaussian clipping noise tends to a Gaussian one after Fourier transformation in optical OFDM receivers \cite{dimitrov2012clip}.

Here, a simulation is performed in order to compare the statistics of the noise added to $\rho_0(\lambda)$ (i.e., direct mapping) and $F_0(t)$ (i.e, GLM-based signaling) after fiber propagation and addition of ASE noise. NFT is applied to the received signal at the output of the fiber, and also an additional IFT operation is performed to obtain the noise on the signal $F_0(t)$. The same simulation parameters as previous sections are used here, and statistics of $\eta_L(\lambda)$ are compared with that of $\Gamma_L (t)$. Note that oversampling ($\kappa > K$) leads to smaller noise variance as a result of averaging effect of Fourier transform operation \cite{safari}. From (\ref{eq_varF}), it can be seen that smaller $h$ (larger number of samples $\kappa$) results in noise reduction. However, in practical scenarios, noise samples become correlated if they are so close to each other as shown in \cite[Section V.D]{imanJLT}. Thus, reducing $h$ is effective up to the point, for which the noise samples are correlated. In this simulation and later in this paper $h=0.01$ is chosen based on observations of noise correlation. In order to investigate the effect of signal-dependency two additional parameters are defined, which represent signal-to-noise ratio (SNR) associated with $\rho(\lambda)$ and $F(t)$ at a specific amplitude of $x$.
\begin{eqnarray}
\mathrm{SNR}_\rho(|x|)=\frac{|x|^2}{\mathsf{E}\{|\eta_L|^2\}|_{|\rho_0|=x}},\\
\mathrm{SNR}_F(|x|)=\frac{|x|^2}{\mathsf{E}\{|\Gamma_L|^2\}|_{|F_0|=x}}.
\end{eqnarray}

\begin{figure}[h]
	\centering
	\begin{subfigure}[t]{0.25\textwidth}
		\centering
		\includegraphics[width=\textwidth]{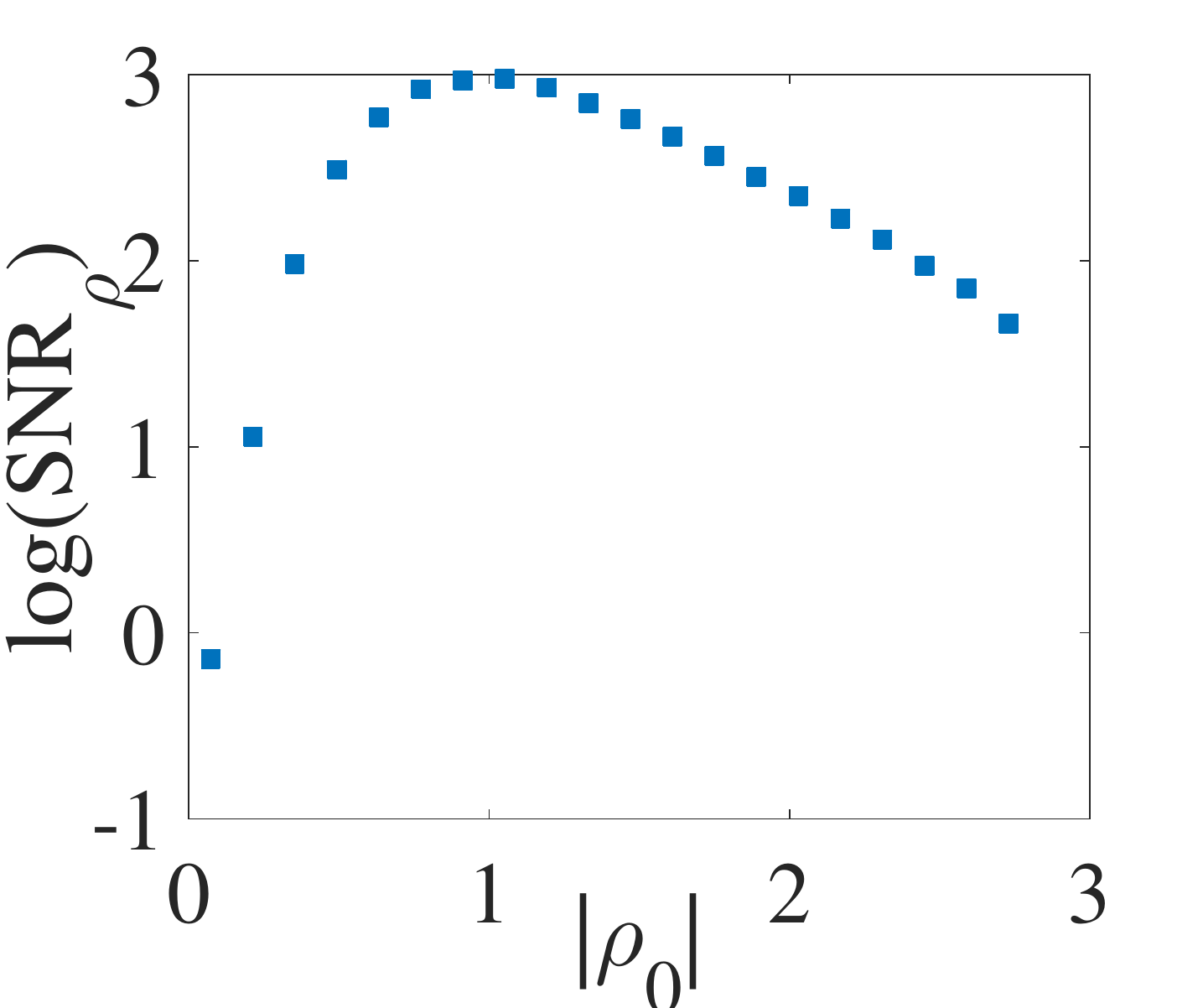} 
		\caption{}
		\label{snr_r}
	\end{subfigure}%
	~
	\begin{subfigure}[t]{0.25\textwidth}
		\centering
		\includegraphics[width=\textwidth]{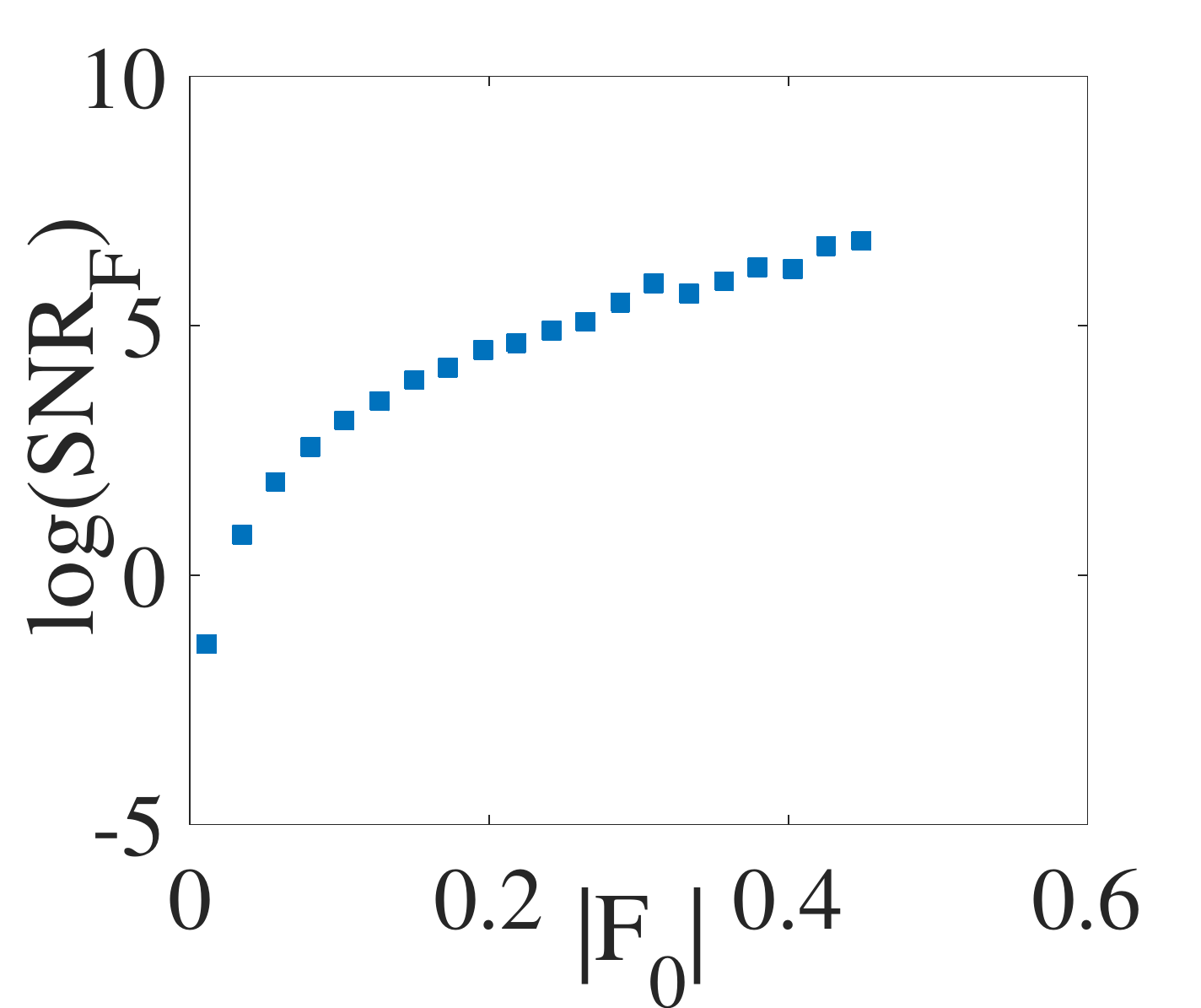} 
		\caption{}
		\label{snr_F}
	\end{subfigure}%
	~
	\begin{subfigure}[t]{0.25\textwidth}         
		\centering
		\includegraphics[width=\textwidth]{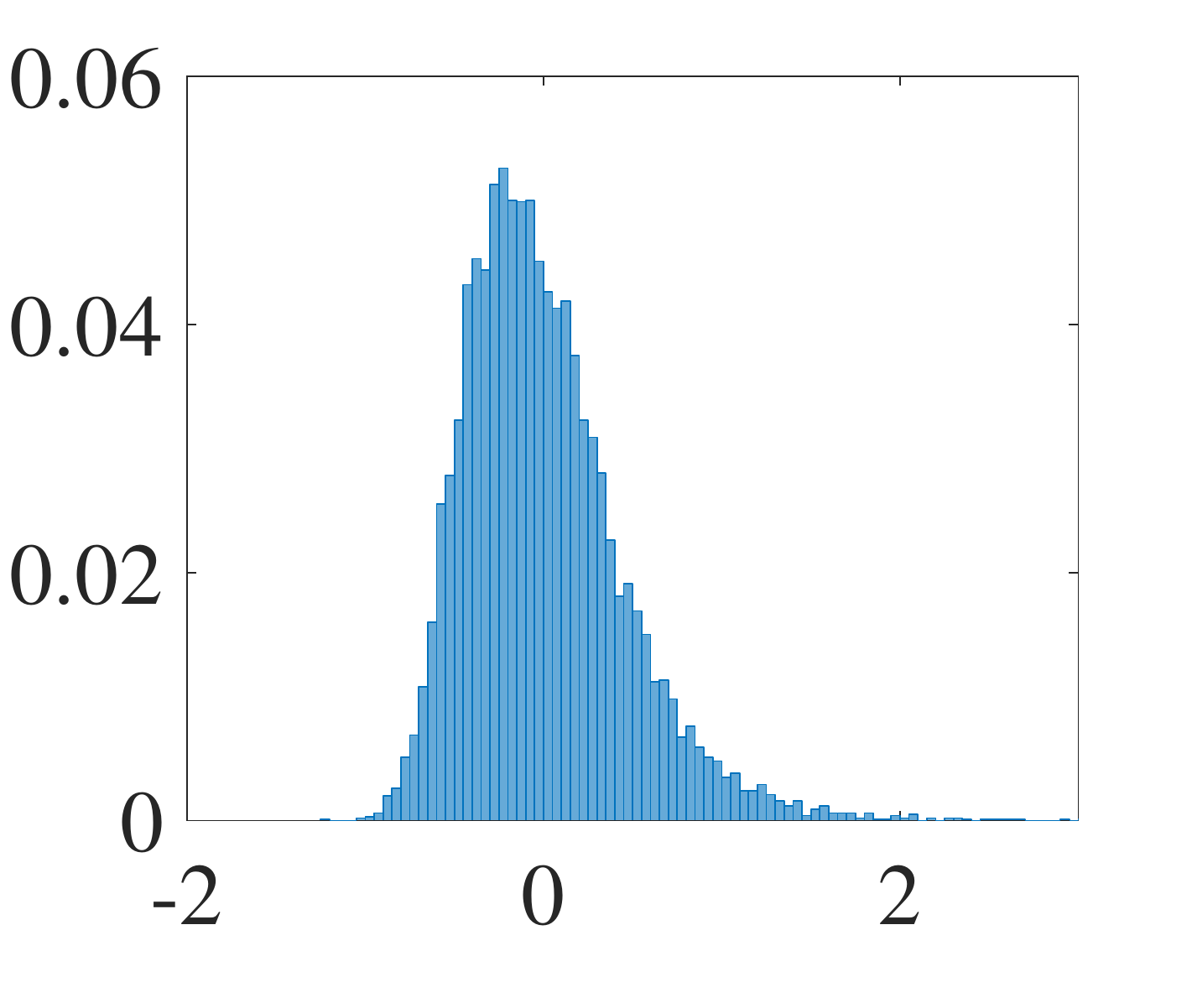}
		\caption{}
		\label{hist_r}
	\end{subfigure}%
	~ 
	\begin{subfigure}[t]{0.25\textwidth}
		\centering
		\includegraphics[width=\textwidth]{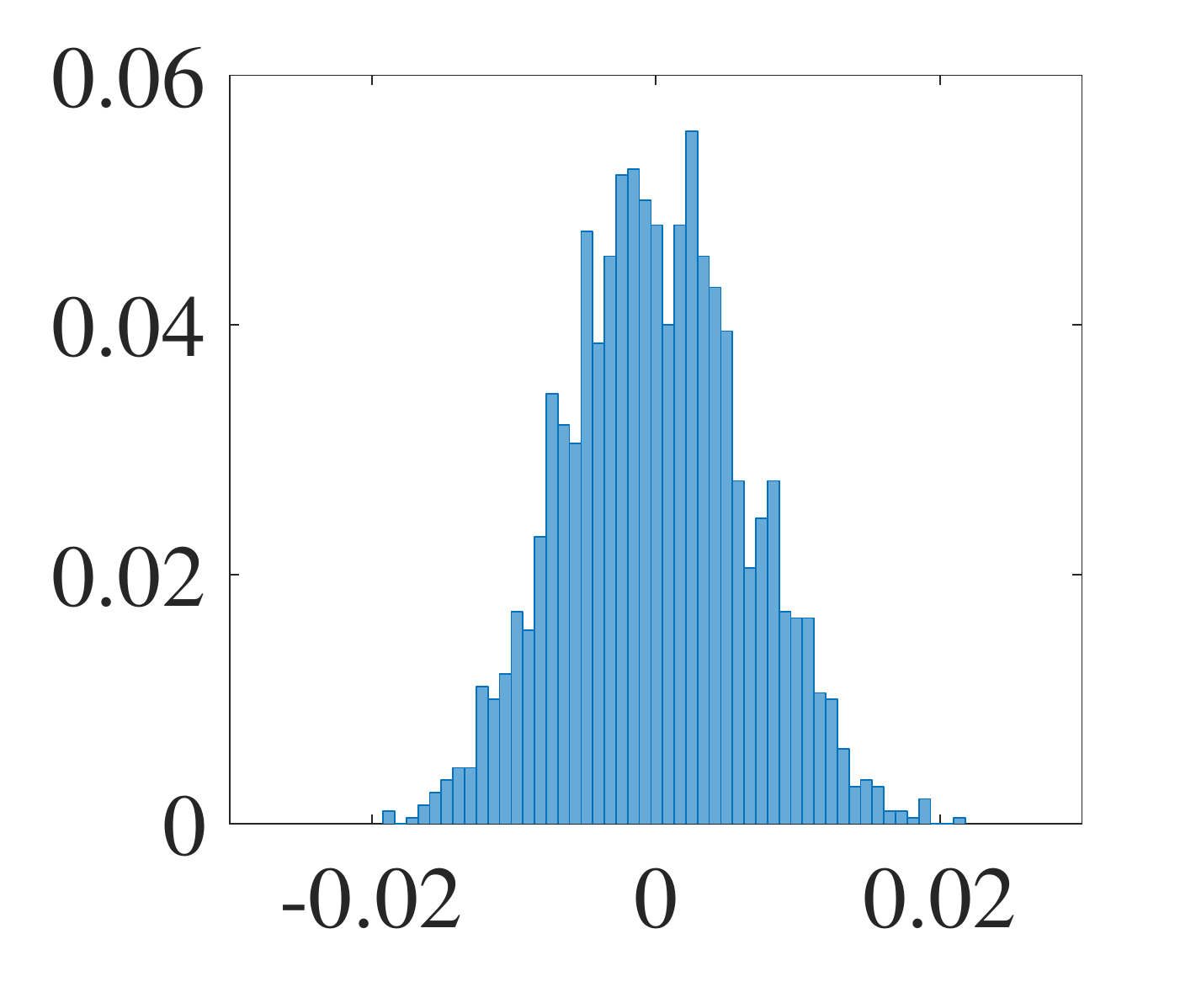} 
		\caption{}
		\label{hist_F}
	\end{subfigure}%
	
	\caption{\small (a) $\mathrm{SNR}_\rho$ for different values of $|\rho_0(\lambda)|$. (b) $\mathrm{SNR}_F$ for different values of $|F_0(t)|$. (c) Distribution of noise $\eta_L(\lambda)$ for $|\rho_0(\lambda)|=2$. (d) Distribution of noise $\Gamma_L(t)$ for all values of $|F_0(t)|$.}
\end{figure}

Figures \ref{snr_r} and \ref{snr_F} show the values for SNR parameters defined above. It can be seen clearly in Figs. \ref{snr_r} and \ref{snr_F} that $\mathrm{SNR}_F$ is increasing while $\mathrm{SNR}_\rho$ has a maximum. Also, the values of $\mathrm{SNR}_F$ are higher than $\mathrm{SNR}_\rho$. Thus, better error performance is expected if data is mapped on $F_0(t)$. Furthermore, the probability distribution for $\Gamma_L(t)$ has a Gaussian-like distribution with constant variance for all signal amplitudes, as shown in Fig. \ref{hist_F}, while the distribution of $\eta_L(\lambda)$ is non-Gaussian, as depicted in Fig. \ref{hist_r}. The signal-independency of noise $\Gamma_L(t)$ will be also shown numerically later in Fig. \ref{scatter_F} of Section \ref{sec_error}. We can therefore conclude that if data is mapped on $F_0(t)$ (i.e. GLM-based signaling), it is distorted by an approximately additive signal-independent Gaussian noise as in the conventional AWGN channels where the underlying uniform amplitude signaling remains efficient. On the other hand, in the direct mapping scheme, the uniform distribution of the symbol amplitudes causes low SNR at higher amplitudes rendering it essentially inefficient at high energy levels.

\section{Numerical results and discussions}
In this section, we present numerical results on the performance of different signaling techniques discussed earlier. Two-ring constellations are used for modulation in order to demonstrate the effect of dependence of noise to the signal amplitude. Unless otherwise stated, $K=128$ symbols are mapped on the nonlinear spectral width of $\Lambda=8$, which are chosen randomly based on the two-ring constellation. First, the methods described in previous section are compared regarding symbol error rate (SER) only taking into account the error due to noise on the amplitude of two particular symbols. Then, bit error rate (BER) performance is investigated based on different constellation formats on two rings. The performance results are presented against the energy of the CS signal denoted by $E$. The channel bandwidth is 26 GHz. 
\subsection{Error rate performance} \label{sec_error}
\begin{figure}[h]
	\centering
	\begin{subfigure}[t]{0.25\textwidth}         
		\centering
		\includegraphics[width=\textwidth]{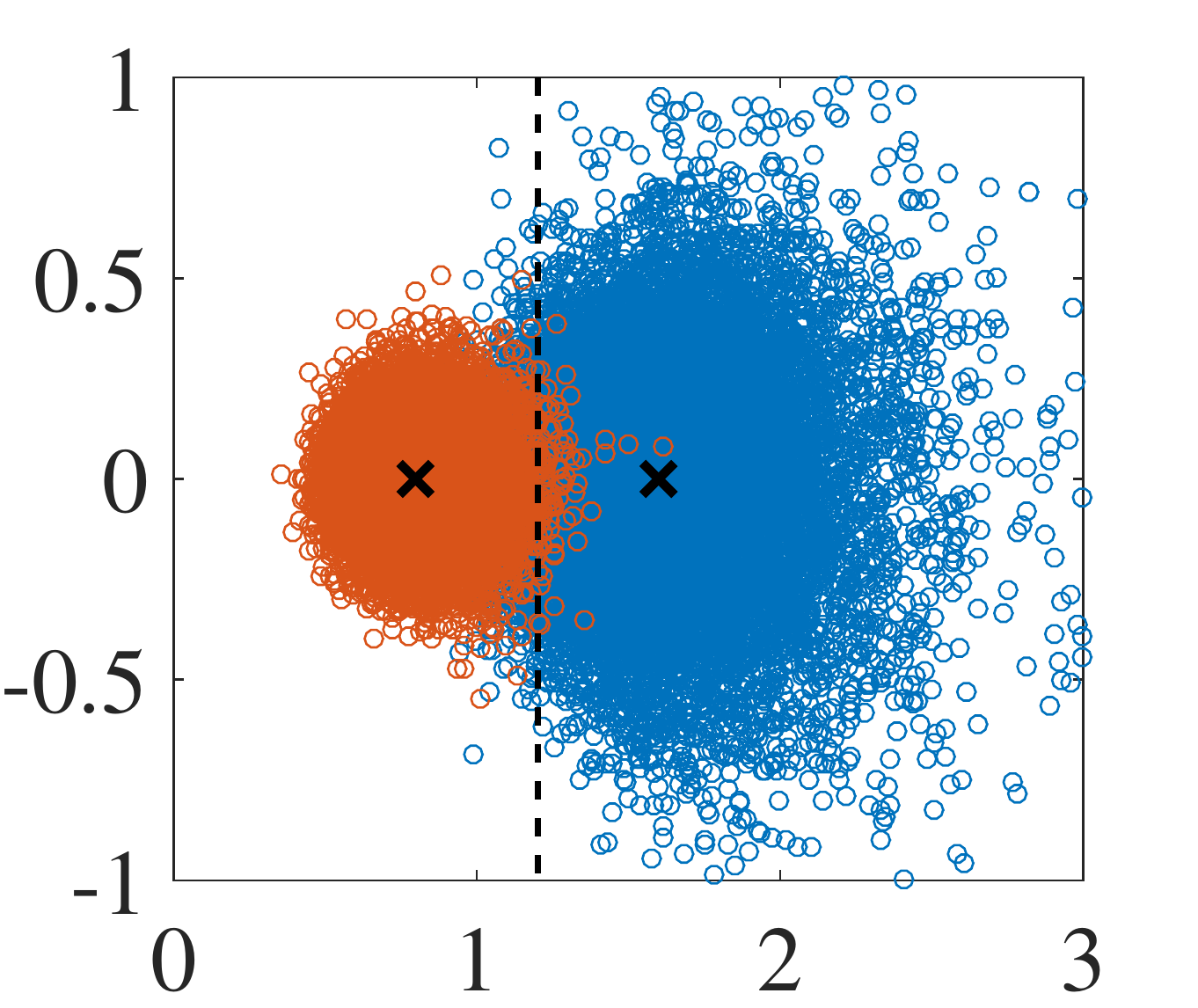}
		\caption{}
		\label{scatter_nft}
	\end{subfigure}%
	~ 
	\begin{subfigure}[t]{0.25\textwidth}
		\centering
		\includegraphics[width=\textwidth]{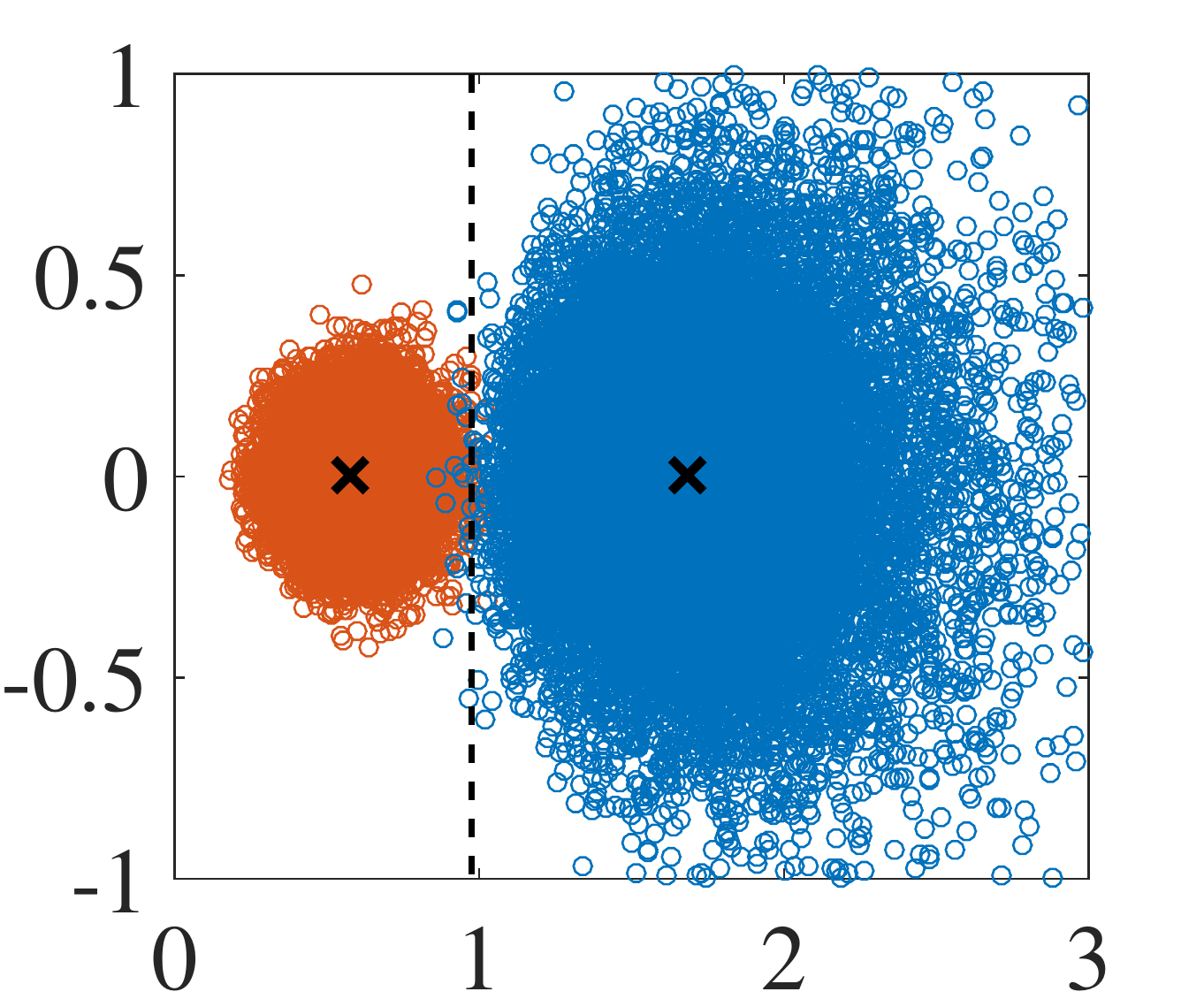} 
		\caption{}
		\label{scatter_vnt}
	\end{subfigure}%
	~
	\begin{subfigure}[t]{0.25\textwidth}
		\centering
		\includegraphics[width=\textwidth]{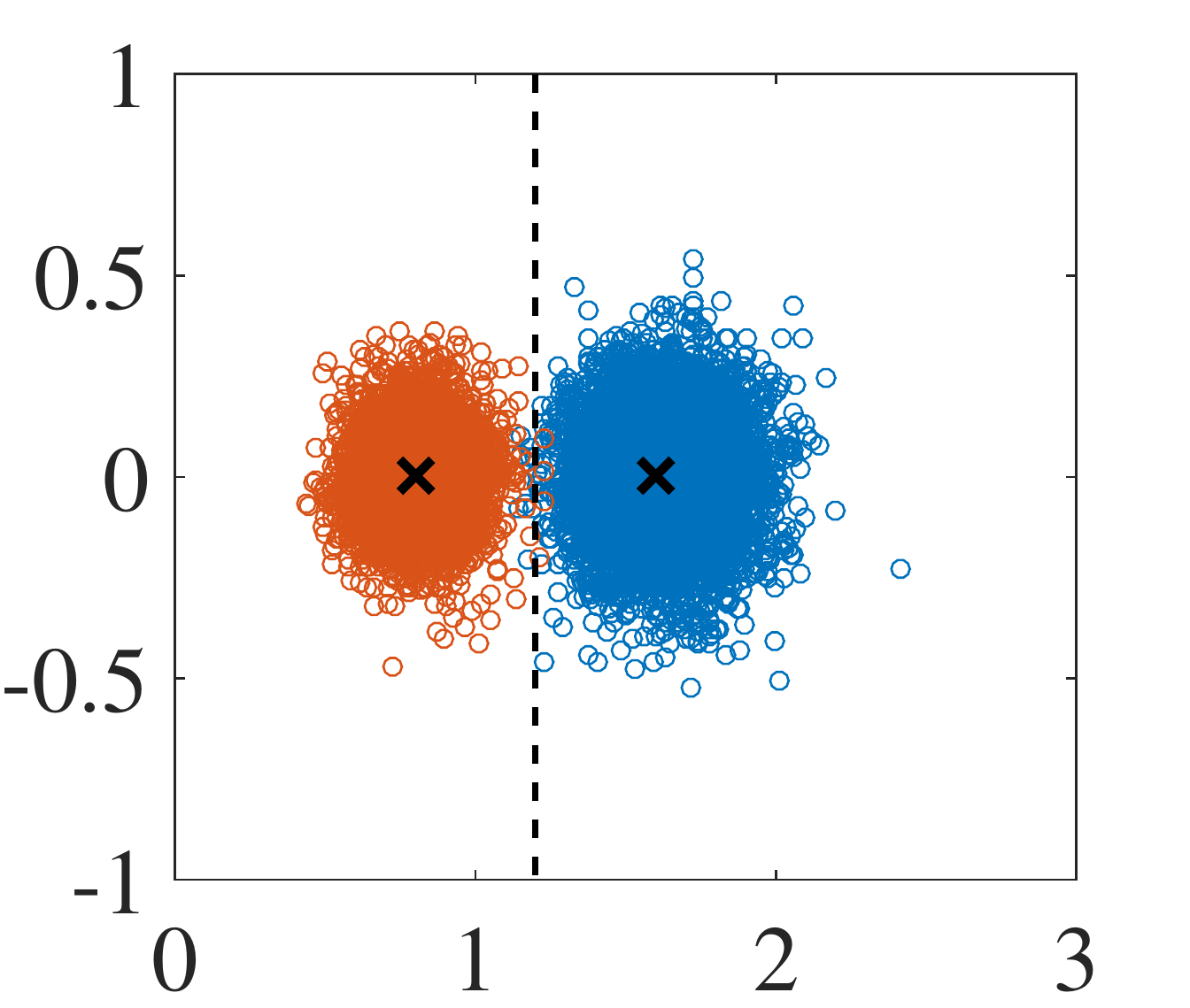} 
		\caption{}
		\label{scatter_fil}
	\end{subfigure}%
	~ 
	\begin{subfigure}[t]{0.25\textwidth}
		\centering
		\includegraphics[width=\textwidth]{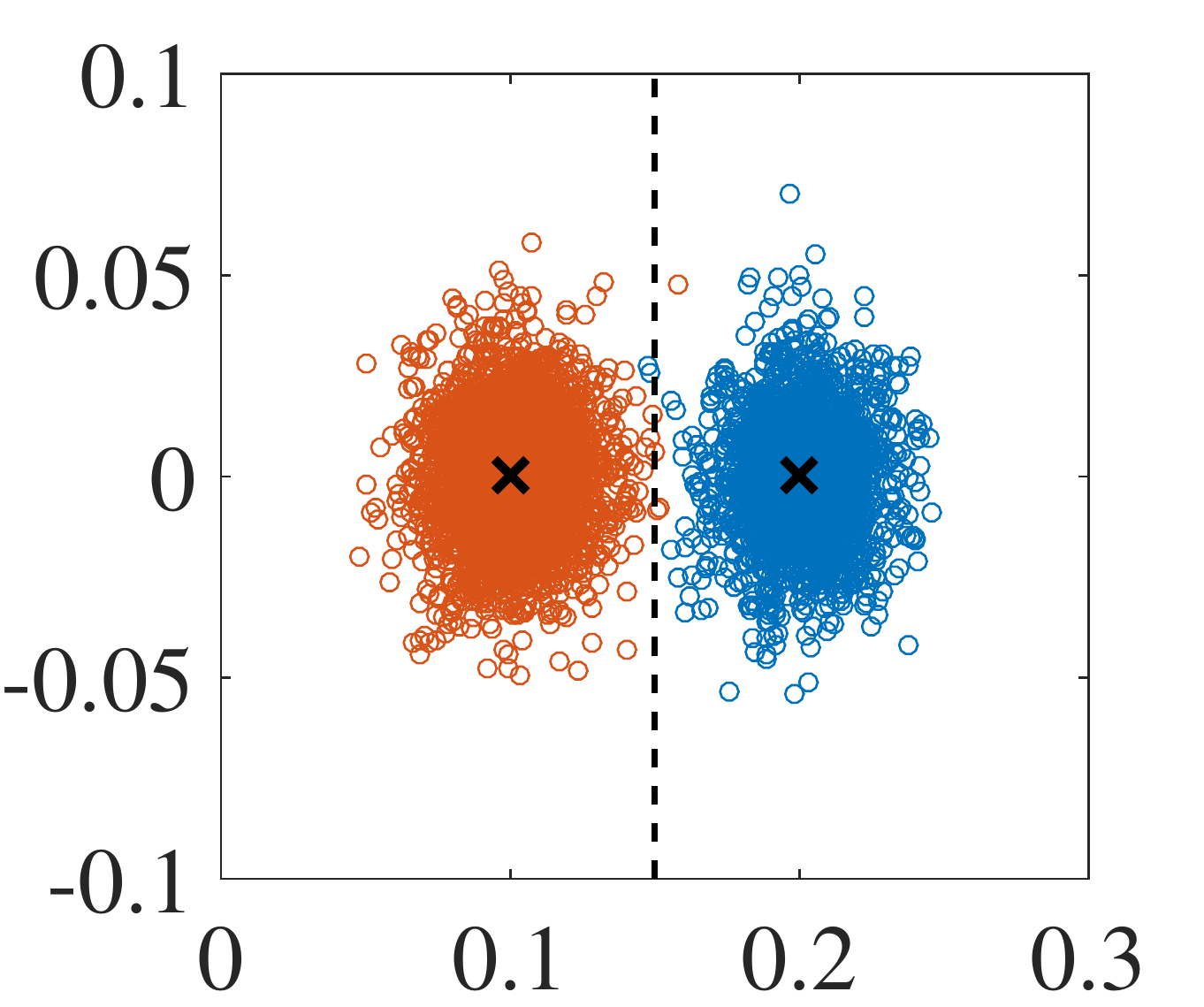} 
		\caption{}
		\label{scatter_F}
	\end{subfigure}%
	\caption{\small Received noisy symbols at $L=2000$ km based on (a) Direct mapping on CS, (b) Nonuniform signaling, (c) Mapping on CS and filtering, and (d) GLM-based signaling.}
	\label{scatter}	
\end{figure}

In the first simulation, SER is compared for different signaling methods assuming only two symbols on the two rings with an equal phase of zero. This allows us to investigate the effect of signal-dependency of noise. Figure \ref{scatter} shows the received noisy symbols at the end of a $L=2000$ km fiber for different signaling methods with $E=2$. Figure \ref{scatter_nft} shows the result for the benchmark scheme, where the symbols are directly mapped on CS as explained in section \ref{sec_model}. It is observed that the variance of both real and imaginary parts (or amplitude and phase) of the noise is dependent on the amplitude of the original transmitted symbol. Figure \ref{scatter_vnt} demonstrates the received symbol for the same energy but for nonuniform levels determined by VNT. In Fig. \ref{scatter_nft}, ring radii are 0.8 and 1.6, and the midpoint decision boundary is at 1.2. However, in Fig. \ref{scatter_vnt}, ring radii are 0.575 and 1.685 with the decision boundary 0.975. Note that the statistics of the noise in CS does not change by applying the VNT method, but the performance improvement is achieved as a results of choosing optimal levels and decision boundary for the rings. For the next two signaling methods in Figs. \ref{scatter_fil} and \ref{scatter_F}, the magnitude of both amplitude and phase of the noise changes. As shown in Fig. \ref{scatter_fil}, when the linear filtering is applied, as described in section \ref{sec_sig_fil}, the noise is significantly reduced. For this simulation, an oversampling of 20 CS samples per symbol is considered. Ring radii and decision boundaries are the same as Fig. \ref{scatter_nft}. It should be noted that the the noise is still signal-dependent although significantly reduced. Figure \ref{scatter_F} depicts, the received symbols when GLM-based signaling is employed. The levels are determined so that the energy $E$ is the same as previous methods. It can be observed that the noise statistics are almost uniform for two levels as predicted in section \ref{sec_sig_F}. Comparing Figs. 8(a) and 8(d), it is observed how crucial it is to whether directly map digital data on CS or on the GLM kernel as the signal would experience two completely different channels. 

\begin{figure}[h]
	\centering
	\includegraphics[width=0.8\textwidth]{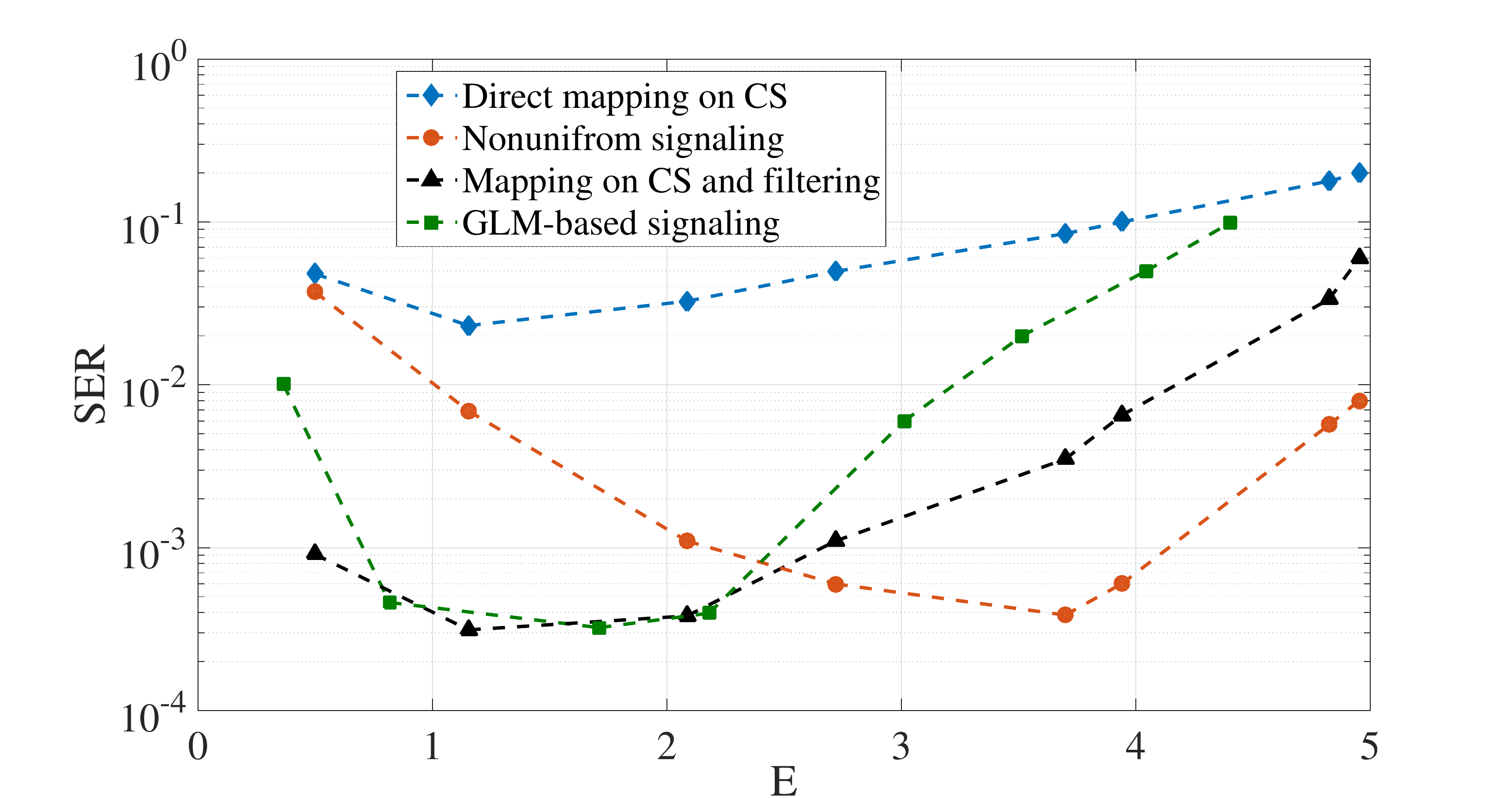} 
	\caption{\small SER for different signaling methods on the CS channel.}
	\label{ser}
\end{figure}

Symbol error rate for two symbols considered in Fig. \ref{scatter} are calculated and shown in Fig. \ref{ser} for different signal energies. It is observed that all proposed methods result in improved performance compared to the direct mapping on CS. The minimum SER is achieved at similar energies for three out of the four methods, but for VNT-based nonuniform signaling the optimum point is different. We conjecture that after $E=3.7$ the VNT is not effective because the noise variance is very large, and the large mean condition for the VNT is not satisfied. For the filtering method, despite noise reduction, the signal-dependency of noise is not eliminated, and thus performance is degraded after a specific energy. For the GLM-based signaling, although signal-dependency is eliminated for a constant energy, the noise variance increases for signals with higher energies based on (\ref{eq_varF}), where higher error rate is expected.
  
\begin{figure}[h]
	\centering
	\begin{subfigure}{0.48\textwidth}
		\centering
		\includegraphics[width=\textwidth]{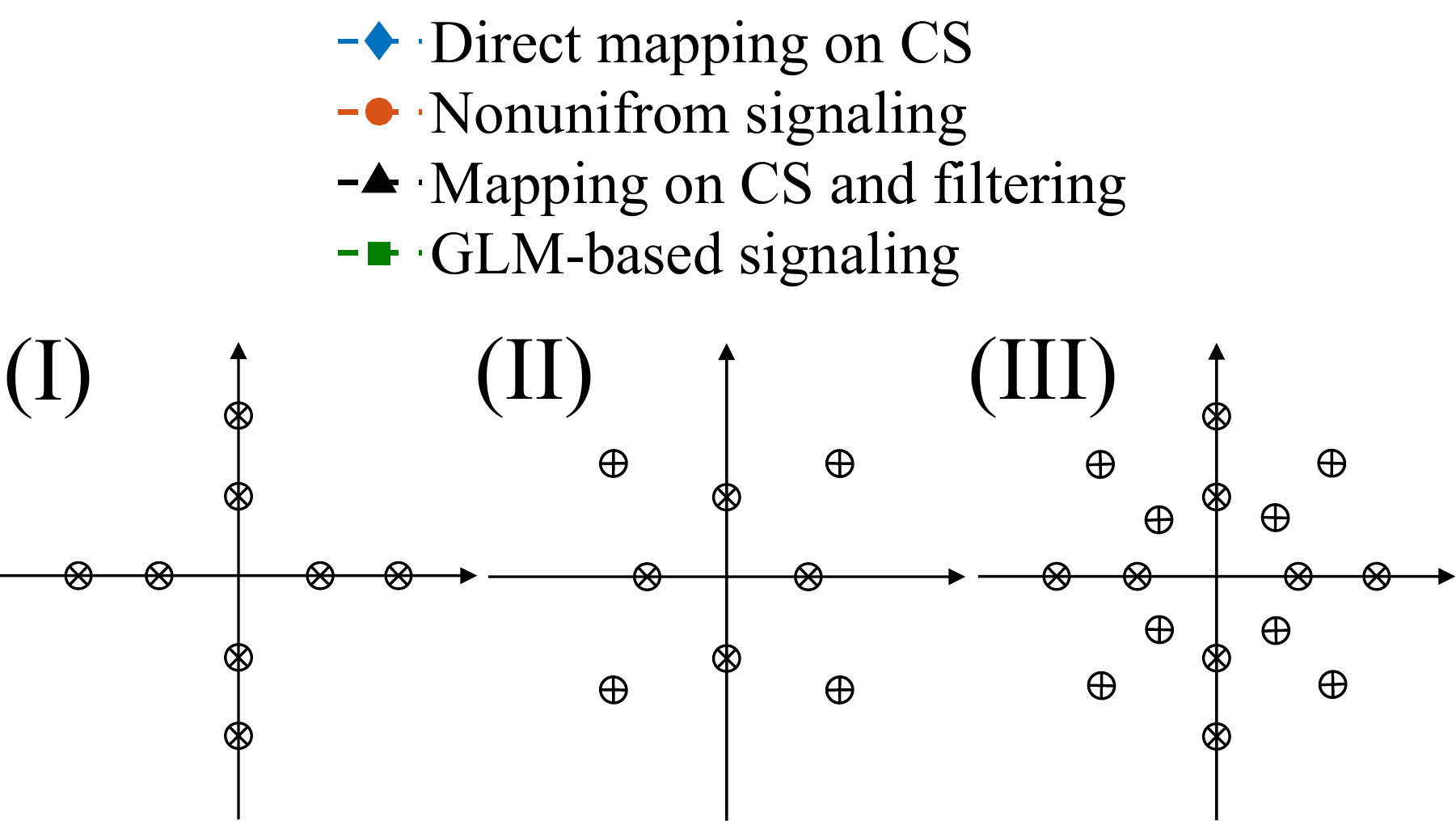} 
		\caption{}
		\label{leg}
	\end{subfigure}
	~
	\begin{subfigure}{0.48\textwidth}
		\centering
		\includegraphics[width=\textwidth]{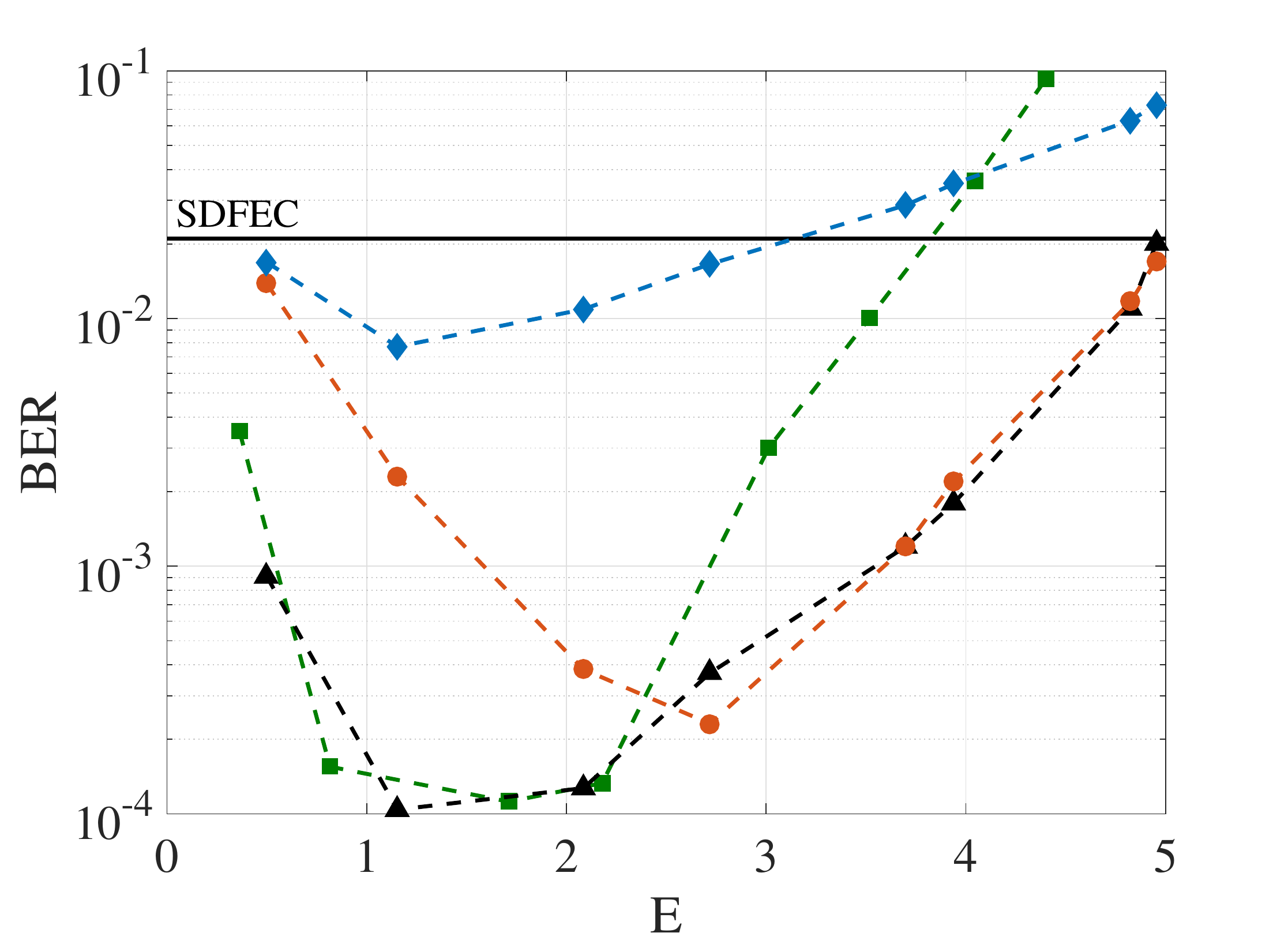} 
		\caption{}
		\label{ber4}
	\end{subfigure}
	\\
	\begin{subfigure}{0.48\textwidth}
			\centering
			\includegraphics[width=\textwidth]{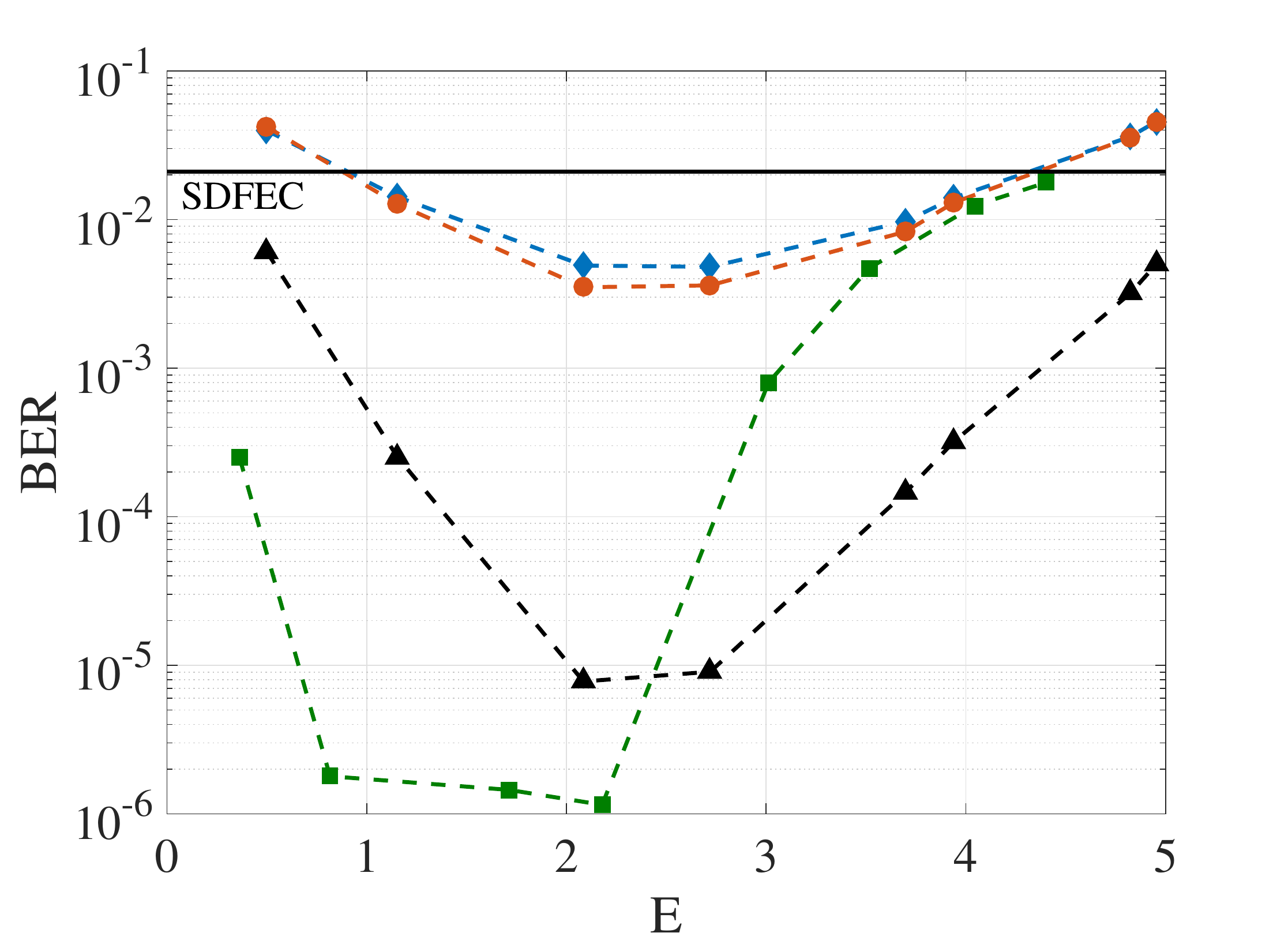} 
			\caption{}
			\label{ber4s}
	\end{subfigure}
	~
	\begin{subfigure}{0.48\textwidth}
		\centering
		\includegraphics[width=\textwidth]{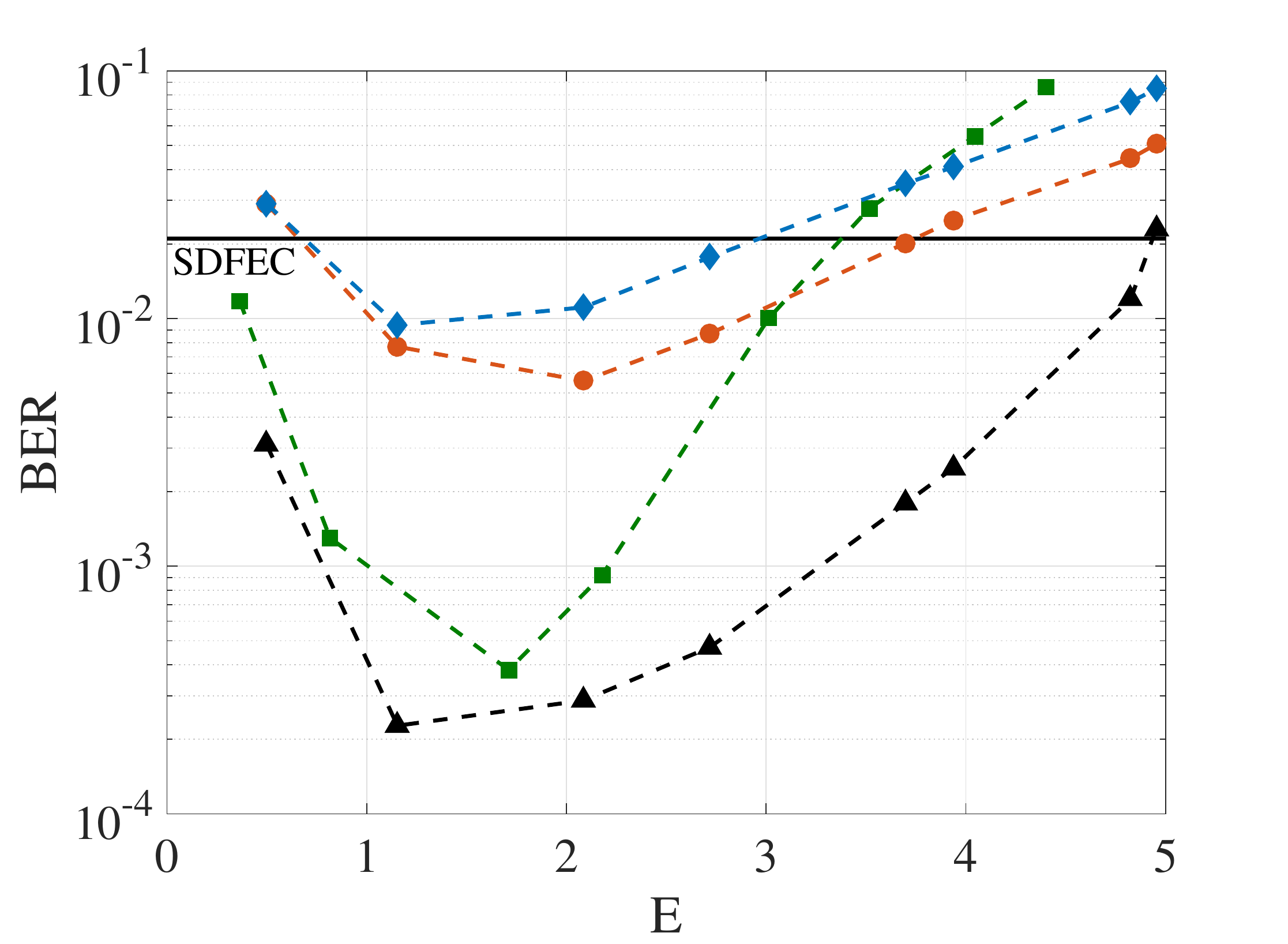} 
		\caption{}
		\label{ber8}
	\end{subfigure}
	\caption{\small (a) Figure legends and constellation diagrams. (b)-(d) BER for different signaling methods at $L=2000$ km for constellation (I)-(III) respectively.}
	\label{ber}
\end{figure}

In Fig. \ref{ber}, the BER performance for three different constellation designs based on the two-ring pattern is depicted. In general, lower BER are achieved compared to direct mapping on CS, when each of the proposed methods is employed, and therefore, higher achievable data rates are expected. However, since NFT-based transmission only eliminates the signal-signal nonlinear interactions, it is expected that signal-noise and noise-noise nonlinear interactions in time domain would eventually limit the performance. Soft decision forward error correction (SDFEC) threshold of $2 \times 10^{-2}$ \cite{wai} is also shown in figures. For constellation (I) in Fig. \ref{ber4}, the minimum BER is achieved with similar values for the filtering method and GLM-based signaling. Since the symbols are well separated in each ring the behavior of BER for different methods is similar to SER in Fig. \ref{ser}. Nevertheless, when the outer ring is $\pi/4$ shifted in phase all the methods results in better BER except the nonuniform signaling based on VNT. This is because the decision boundary levels for nonuniform signaling are defined based on the one dimensional (i.e., real) signal space considered in Section \ref{sec_sig_vnt}, which are not optimum for the complex constellation (II) in Fig. \ref{ber4s}. An optimal set of decision boundaries could be only determined based on maximum likelihood requiring full analytical description of the noise statistics, which are not available or through exhaustive numerical search. Note that, unlike constellation (II), the decision boundary levels defined based on one dimensional signal space are nearly optimal for constellation (I) since errors dominantly occur in the radial directions.

In Fig. \ref{ber8}, 8 symbols are chosen on each ring, and BER is obtained by simulation. For constellation (III), error originated from the noise on the phase is dominant, which causes the BER for nonuniform signaling not to be as low as other proposed methods. For the GLM-based signaling, the slope of the performance degradation is higher compared to all other methods. This can be explained by referring to equation (\ref{eq_varF}). Based on this equation the noise variance increases for higher signal energies with a polynomial of order 4, and this happens equally for all symbol amplitudes. However, for other methods, the inner ring always experiences lower error rate compared to the outer ring resulting in a slower increase of BER on average. It should be noted that, as explained in Section \ref{sec_disp}, the method proposed in \cite{imanCLEO} can be used to further improve the performance.

Note that, in Fig. \ref{ber}, operating bandwidth is constant and BER is obtained for various energies, but the effective data rates vary for different signaling methods and constellations. In fact, the required temporal window for NFT operation without numerical error is different for each signal energy $E$ and signaling methods. This happens because of the effect of residual tail at the output of INFT and the different effect of dispersion. For instance, at $E=2$, taking into account the dispersion effect, for Figs. \ref{ber4} and \ref{ber4s} the effective bit rate is 21.9 Gpbs for first three methods and 20.7 Gbps for GLM-based signaling, while for constellation in Fig. \ref{ber8} these values are respectively 29.2 and 27.6 Gpbs.

The corresponding launch power levels in Figs. \ref{ser} and \ref{ber} range from -17 dBm to -10dBm. Not that, due to optimization performed for $K=128$ in Section 3, the dispersion effect is significantly reduced and in turn the nonlinear effect is intensified such that the CS channel with power levels above can reach the high nonlinearity regimes where the BER performances start declining. Assuming a larger $K$, the optimal bandwidth increases, which leads to higher launch power; however, studying such a scenario would also require dealing with significantly more time-consuming simulation of NFT and INFT operations.

It is worth mentioning that better performance for techniques presented here are obtained with a cost of computational complexity. In order to compare the complexity of different methods, the number of NFT operations at the receiver are taken into account. For direct mapping on CS and nonuniform signaling, the number of NFT operations is equal to the number of symbols, which, in our case, is $K=128$. For GLM-based signaling, $6\times K$ samples of $\tilde{\rho}_L(\lambda)$ (corresponding to $h=0.01$) are computed (768 NFT operations) for calculation of $F_L(t)$. Also for the linear filtering method, as stated earlier, $20 \times K$ samples of $\tilde{\rho}_L(\lambda)$ are computed (2560 NFT operations), and also additional FFT and IFFT operations are performed.

\subsection{Distance reach}
Considering soft and hard decision FEC thresholds the distance reach are compared for different methods in Fig. 11. For all methods shifted constellation (II) in Fig. \ref{ber} is used except for the nonuniform signaling based on VNT, for which constellation (I) is selected. The input energy $E$ is chosen for the minimum achieved BER for each technique. Considering SDFEC and ignoring the redundant bits due to FEC coding, the simulation results demonstrate that distances as long as 7100 km can be reached by applying the linear filtering technique. The effective bit rate at distance 7100 km is 9.6 Gbps. At second place, with GLM-based signaling, a distance reach of about 5900 km with effective bit rate 11.7 Gpbs is obtained. When the levels are optimized using VNT, the achieved reach  and effective bit rate are 5000 km and 12.7 Gbps. The corresponding values are respectively 3000 km and 18.5 Gpbs for direct mapping on CS. Moreover, considering hard decision forward error correction code (HDFEC) with 7\% overhead with threshold of $3.8\times 10^{-3}$, distance reaches of 1900, 3300, 4300, and 5300 are achieved respectively for direct mapping on CS, nonuniform signaling, GLM-based signaling, and filtering. It should be noted that, for simplicity and better presentation, $E$ and $\Lambda$ are considered to be fixed for all fiber lengths and are chosen based on the best performance of a 2000 km link. However, for further performance improvement, $E$ and $\Lambda$ should be optimized for each individual fiber length. This analysis requires extensive numerical simulation and is out of scope of the current paper as marginal improvement is expected.

\begin{figure}[h]
	\centering
	\includegraphics[width=0.8\textwidth]{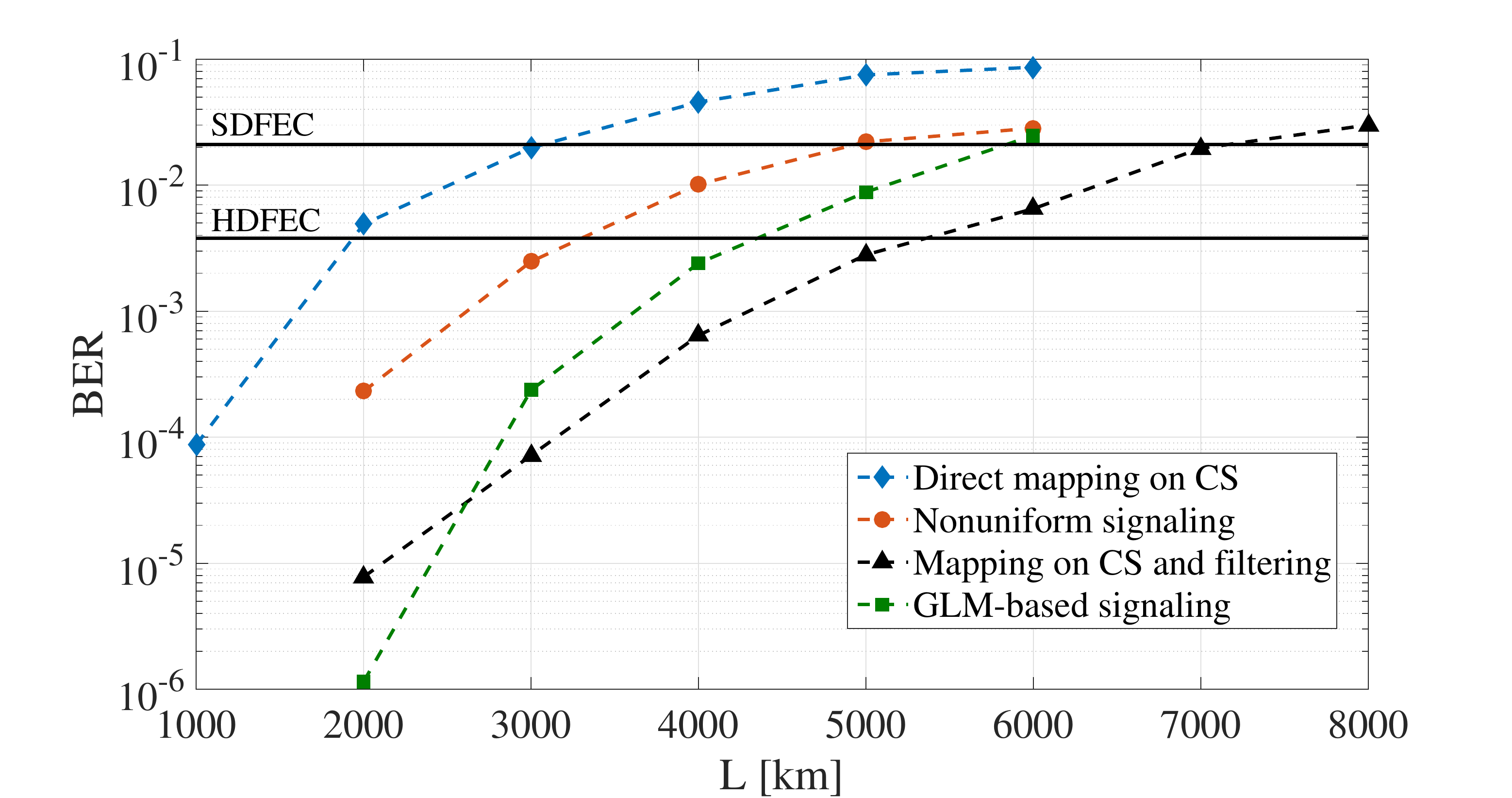} 
	\caption{\small Distance reach for different methods.}
	\label{reach}
\end{figure}
\section{Summary}
Efficient signaling on CS was studied in this paper. First, the effect of chromatic dispersion on NFDM systems was studied. It was demonstrated that the signal bandwidth can be chosen so that the temporal signal broadening at the end of the fiber is minimized. By correctly selecting the bandwidth, not only the data rate is maximized, but also the noise variance is minimized in nonlinear spectral domain. Moreover, three different techniques were investigated for improving error performance compared to direct mapping on CS. For instance, our analysis showed that applying a linear filter on CS at the receiver significantly reduces the noise and improves the BER performance so that 9.6 Gbps can be transfered over 7100 km of fiber by only mapping the data on CS. The signal bandwidth was considered 26 GHz in our numerical results, however, data rates beyond values presented here may potentially be achieved by exploiting higher bandwidth and mapping data on DS as well.
\section*{Acknowledgment}
The authors would like to thank Professor Frank R. Kschischang for insightful discussions on simulation and signaling of the continuous spectrum channel.  


\begin{thebibliography}{10}
	\newcommand{\enquote}[1]{``#1''}
	
	\bibitem{Cartledge}
	J.~C. Cartledge, F.~P. Guiomar, F.~R. Kschischang, G.~Liga, and M.~P. Yankov,
	\enquote{Digital signal processing for fiber nonlinearities,} Opt. Express
	\textbf{25}(3), 1916--1936 (2017).
	
	\bibitem{capacity}
	R.~Essiambre, G.~Kramer, P.~J. Winzer, G.~J. Foschini, and B.~Goebel,
	\enquote{Capacity limits of optical fiber networks,} IEEE/OSA J. Lightw. Technol. \textbf{28}(4), 662--701 (2010).
	
	\bibitem{winzer}
	P.~J. Winzer and D.~T. Neilson, \enquote{From scaling disparities to integrated
		parallelism: A decathlon for a decade,} IEEE/OSA J. Lightw. Technol. \textbf{35}(5), 1099--1115
	(2017).
	
	\bibitem{yousefi1}
	M.~Yousefi and F.~Kschischang, \enquote{Information transmission using the	nonlinear Fourier transform, part I: Mathematical tools,} IEEE Trans. Inform. Theory \textbf{60}(7), 4312--4328 (2014).
	
	\bibitem{yousefi2}
	M.~Yousefi and F.~Kschischang, \enquote{Information transmission using the nonlinear Fourier transform, part II: Numerical methods,} IEEE Trans. Inform. Theory \textbf{60}(7), 4329--4345 (2014).
	
	\bibitem{yousefi3}
	M.~Yousefi and F.~Kschischang, \enquote{Information transmission using the	nonlinear Fourier transform, part III: Spectrum modulation,} IEEE Trans. Inform. Theory \textbf{60}(7), 4346--4369 (2014).
	
	\bibitem{Turitsynoptica}
	S.~K. Turitsyn, J.~E. Prilepsky, S.~T. Le, S.~Wahls, L.~L. Frumin, M.~Kamalian,
	and S.~A. Derevyanko, \enquote{Nonlinear Fourier transform for optical data
		processing and transmission: Advances and perspectives,} Optica \textbf{4}(3),
	307--322 (2017).
	
	\bibitem{prilepsky2014}
	J.~E. Prilepsky, S.~A. Derevyanko, K.~J. Blow, I.~Gabitov, and S.~K. Turitsyn,
	\enquote{Nonlinear inverse synthesis and eigenvalue division multiplexing in
		optical fiber channels,} Phys. Rev. Lett. \textbf{113}(1), 013901 (2014).
	
	\bibitem{yousefi2016nonlinear}
	M.~I. Yousefi and X.~Yangzhang, \enquote{Nonlinear frequency-division
		multiplexing,} arXiv preprint arXiv:1603.04389  (2016).
	
	\bibitem{iman}
	I.~Tavakkolnia and M.~Safari, \enquote{Signalling over nonlinear fibre-optic
		channels by utilizing both solitonic and radiative spectra,} in
	\emph{European Conference on Networks and Communications (EuCNC)}
	(2015), pp. 103--107.
	
	\bibitem{hari2016}
	S.~Hari, M.~I. Yousefi, and F.~R. Kschischang, \enquote{Multieigenvalue
		communication,} IEEE/OSA J. Lightw. Technol. \textbf{34}(13), 3110--3117
	(2016).
	
	\bibitem{haribi}
	S.~Hari and F.~R. Kschischang, \enquote{Bi-directional algorithm for computing
		discrete spectral amplitudes in the NFT,} IEEE/OSA J. Lightw. Technol. \textbf{34}(15), 3529--3537 (2016).
	
	\bibitem{buelow7}
	H.~Buelow, V.~Aref, and W.~Idler, \enquote{Transmission of waveforms determined
		by 7 eigenvalues with PSK-modulated spectral amplitudes,} in \emph{European Conference on Optical Communication}  (2016), pp. 1--3.
	
	\bibitem{bulow2015experimental}
	H.~B{\"u}low, \enquote{Experimental demonstration of optical signal detection
		using nonlinear Fourier transform,} IEEE/OSA J. Lightw. Technol.
	\textbf{33}(7), 1433--1439 (2015).
	
	\bibitem{dong2015nonlinear}
	Z.~Dong, S.~Hari, T.~Gui, K.~Zhong, M.~I. Yousefi, C.~Lu, P.-K.~A. Wai, F.~R.
	Kschischang, and A.~P.~T. Lau, \enquote{Nonlinear frequency division
		multiplexed transmissions based on NFT,} IEEE Photon. Technol. Lett.
	\textbf{27}(15), 1621--1623 (2015).

	\bibitem{wai}
	T.~Gui, T.~H. Chan, C.~Lu, A.~P.~T. Lau, and P.~K.~A. Wai, \enquote{Alternative
		decoding methods for optical communications based on nonlinear Fourier
		transform,} IEEE/OSA J. Lightw. Technol. \textbf{35}(9),1542--1550 (2017).
	
	\bibitem{shevchenko2016lower}
	N.~A. Shevchenko, S.~A. Derevyanko, J.~E. Prilepsky, A.~Alvarado, P.~Bayvel,
	and S.~K. Turitsyn, \enquote{On the capacity of the noncentral Chi-channel
		with applications to soliton amplitude modulation,} arXiv preprint
	arXiv:1609.02318  (2016).
	
	\bibitem{alex}
	N.~A. Shevchenko, J.~E. Prilepsky, S.~A. Derevyanko, A.~Alvarado, P.~Bayvel,
	and S.~K. Turitsyn, \enquote{A lower bound on the per soliton capacity of the
		nonlinear optical fibre channel,} in \emph{Information Theory
		Workshop}  (2015), pp. 104--108.
	
	\bibitem{aref2016demonstration}
	V.~Aref, S.~T. Le, and H.~Buelow, \enquote{Demonstration of fully nonlinear
		spectrum modulated system in the highly nonlinear optical transmission
		regime,} in \emph{European Conference on Optical Communication}  (2016), pp. 1--3.
	
	\bibitem{wahls2015fast}
	S.~Wahls and H.~V. Poor, \enquote{Fast numerical nonlinear Fourier transforms,} IEEE Trans. Inform. Theory \textbf{61}(12), 6957--6974 (2015).
	
	\bibitem{wahls2016fast}
	S.~Wahls and V.~Vaibhav, \enquote{Fast inverse nonlinear Fourier transforms for
		continuous spectra of Zakharov-Shabat type,} arXiv preprint arXiv:1607.01305
	(2016).
	
	\bibitem{civelli2015numerical}
	S.~Civelli, L.~Barletti, and M.~Secondini, \enquote{Numerical methods for the
		inverse nonlinear Fourier transform,} in \emph{International Workshop on Digital Communications}  (2015), pp.	13--16.
	
	\bibitem{kamalian2016periodic1}
	M.~Kamalian, J.~E. Prilepsky, S.~T. Le, and S.~K. Turitsyn, \enquote{Periodic
		nonlinear Fourier transform for fiber-optic communications, part F: theory
		and numerical methods,} Opt. Express \textbf{24}(16), 18353--18369 (2016).
	
	\bibitem{kamalian2016periodic2}
	M.~Kamalian, J.~E. Prilepsky, S.~T. Le, and S.~K. Turitsyn, \enquote{Periodic
		nonlinear Fourier transform for fiber-optic communications, part II:
		eigenvalue communication,} Opt. Express \textbf{24}(16), 18370--18381 (2016).
	
	\bibitem{nature}
	S.~A. Derevyanko, J.~E. Prilepsky, and S.~K. Turitsyn, \enquote{Capacity
		estimates for optical transmission based on the nonlinear Fourier transform,}
	Nature Commun. \textbf{7}, 12710 (2016).
	
	\bibitem{imanJLT}
	I.~Tavakkolnia and M.~Safari, \enquote{Capacity analysis for the continuous
		spectrum of nonlinear optical fiber,} IEEE/OSA J. Lightw. Technol.
	\textbf{35}(11), 2086--2097 (2017).

	\bibitem{yangzhang2016nonlinear}
	X.~Yangzhang, M.~I. Yousefi, A.~Alvarado, D.~Lavery, and P.~Bayvel,
	\enquote{Nonlinear frequency-division multiplexing in the focusing regime,}
	in \emph{Optical Fiber Communication Conference} (2017), p. Tu3D-1.
	
	\bibitem{le2014nonlinear}
	S.~T. Le, J.~E. Prilepsky, and S.~K. Turitsyn, \enquote{Nonlinear inverse
		synthesis for high spectral efficiency transmission in optical fibers,}
	Opt. Express \textbf{22}(22), 26720--26741 (2014).
	
	\bibitem{le2015nonlinear}
	S.~T. Le, J.~E. Prilepsky, and S.~K. Turitsyn, \enquote{Nonlinear inverse
		synthesis technique for optical links with lumped amplification,} Opt.
	Express \textbf{23}(7), 8317--8328 (2015).
	
	\bibitem{le2016demonstration}
	S.~T. Le, I.~D. Philips, J.~E. Prilepsky, P.~Harper, A.~D. Ellis, and S.~K.
	Turitsyn, \enquote{Demonstration of nonlinear inverse synthesis transmission
		over transoceanic distances,} IEEE/OSA J. Lightw. Technol. \textbf{34}(10),
	2459--2466 (2016).
	
	\bibitem{le2016modified}
	S.~T. Le, J.~E. Prilepsky, P.~Rosa, J.~D. Ania-Castanon, and S.~K. Turitsyn,
	\enquote{Nonlinear inverse synthesis for optical links with distributed Raman
		amplification,} IEEE/OSA J. Lightw. Technol. \textbf{34}(8), 1778--1786
	(2016).
	
	\bibitem{imanCLEO}
	I.~Tavakkolnia and M.~Safari, \enquote{Dispersion pre-compensation for
		NFT-based optical fiber communication systems,} in \emph{Conference on
		Lasers and Electro-Optics} (2016), p. SM4F4.
	
	\bibitem{ablowitz}
	M.~J. Ablowitz, B.~Prinari, and A.~D. Trubatch, \emph{Discrete and continuous
		nonlinear Schr{\"o}dinger systems} (Cambridge University, 2004).
	
	\bibitem{agrawal0}
	G.~P. Agrawal, \emph{Nonlinear fiber optics} (Academic, 2013), 5th ed.
	
	\bibitem{sorokina2016ripple}
	M.~Sorokina, S.~Sygletos, and S.~Turitsyn, \enquote{Ripple distribution for
		nonlinear fiber-optic channels,} Opt. Express \textbf{25}(3), 2228--2238 (2017).
	
	\bibitem{skidin2016mitigation}
	A.~S. Skidin, O.~S. Sidelnikov, M.~P. Fedoruk, and S.~K. Turitsyn,
	\enquote{Mitigation of nonlinear transmission effects for OFDM 16-QAM optical
		signal using adaptive modulation,} Opt. Express \textbf{24}(26), 30296--30308
	(2016).
	
	\bibitem{steve}
	A.~A. Farid and S.~Hranilovic, \enquote{Channel capacity and non-uniform
		signalling for free-space optical intensity channels,} IEEE J. Sel. Areas Commun. \textbf{27}(9), 1553--1563 (2009).
	
	\bibitem{bartlett}
	M.~Bartlett, \enquote{The square root transformation in analysis of variance,}
	Supplement to the Journal of the Royal Statistical Society pp. 68--78 (1936).
	
	\bibitem{curtiss}
	J.~H. Curtiss, \enquote{On transformations used in the analysis of variance,}
	The Annals of Mathematical Statistics \textbf{14}, 107--122 (1943).
	
	\bibitem{safari}
	M.~Safari, \enquote{Efficient optical wireless communication in the presence of
		signal-dependent noise,} in \emph{International Conference on Communication Workshop (ICCW)}  (2015), pp. 1387--1391.
	
	\bibitem{tsiatmas}
	A.~Tsiatmas, F.~M. Willems, and C.~P. Baggen, \enquote{Square root
		approximation to the Poisson channel,} in \emph{International Symposium on Information Theory}  (2013), pp.
	1695--1699.
	
	\bibitem{imanROF}
	I.~Tavakkolnia and M.~Safari, \enquote{Variance normalizing transform for
		performance improvement in radio-over-fiber systems,} in \emph{European Conference on Networks
		and Communications} (2016), pp. 255-259.
	
	\bibitem{prucnal}
	P.~R. Prucnal and B.~E. Saleh, \enquote{Transformation of
		image-signal-dependent noise into image-signal-independent noise,} Opt.
	lett. \textbf{6}(7), 316--318 (1981).
	
	\bibitem{ofdm}
	J.~Rice, \emph{Mathematical Statistics and Data Analysis} (Duxbury, 1995), 2nd
	ed.
	
	\bibitem{imaneff}
	 I.~Tavakkolnia and M.~Safari, \enquote{Efficient Signalling on the Continuous Spectrum of Nonlinear Optical Fibre,} in \emph{Conference on Lasers and Electro-Optics/Europe and the European Quantum Electronics Conference}  (2017).
 	
	\bibitem{frumin2017new}
	L.~L. Frumin, A.~A. Gelash, and S.~K. Turitsyn, \enquote{New approaches to coding information using inverse scattering transform, } Phys. Rev. Lett. \textbf{118}(22), 223901 (2017).
	
	\bibitem{dimitrov2012clip}
	S.~Dimitrov, S.~Sinanovic, and H.~Haas, \enquote{Clipping Noise in OFDM-Based Optical Wireless Communication Systems,} IEEE Trans. Commun. \textbf{60} (4), 1072--1081 (2012).  
	
\end{thebibliography}
\end{document}